\documentclass{aa}
\usepackage{graphicx}
\usepackage{url}

\usepackage{txfonts}
\usepackage[breaklinks,colorlinks,citecolor=blue]{hyperref}
\usepackage{caption}
\usepackage{longtable}
\usepackage{multicol}
\usepackage{natbib,twoopt}
\usepackage{soul}

\usepackage[switch]{lineno}

\newcommand{\GG}[1]{}

\bibpunct{(}{)}{;}{a}{}{,} %% natbib format for A&A and ApJ
\makeatletter
\newcommandtwoopt{\citeads}[3][][]{\href{http://adsabs.harvard.edu/abs/#3}%
{\def\hyper@linkstart##1##2{}%
\let\hyper@linkend\@empty\citealp[#1][#2]{#3}}}
\newcommandtwoopt{\citepads}[3][][]{\href{http://adsabs.harvard.edu/abs/#3}%
{\def\hyper@linkstart##1##2{}%
\let\hyper@linkend\@empty\citep[#1][#2]{#3}}}
\newcommandtwoopt{\citetads}[3][][]{\href{http://adsabs.harvard.edu/abs/#3}%
{\def\hyper@linkstart##1##2{}%
\let\hyper@linkend\@empty\citet[#1][#2]{#3}}}
\newcommandtwoopt{\citeyearads}[3][][]%
{\href{http://adsabs.harvard.edu/abs/#3}
{\def\hyper@linkstart##1##2{}%
\let\hyper@linkend\@empty\citeyear[#1][#2]{#3}}}
\makeatother

\newcommand{\kms}{km~s$^{-1}$}

\newcommand{\sqdeg}{deg$^2$}

\def \deg         {\text{$^{\circ}$}}

\def \arcsec      {\text{$^{\prime\prime}$}}

\def \mujybeam    {$\mathrm{\mu}$Jy\,beam$^{-1}$}
\begin{document} 
% \linenumbers

\title{Search and analysis of giant radio galaxies with associated nuclei (SAGAN).}

\subtitle{V. Study of giant double-double radio galaxies from LoTSS DR2}
   \titlerunning{SAGAN-V: Giant DDRGs from LoTSS DR2}
 % \subtitle{}

\author {Pratik Dabhade\inst{1,2,3}\thanks{E-mail: pratik.dabhade@ncbj.gov.pl, pratikdabhade13@gmail.com}
\and Kshitij Chavan\inst{4}
\and D.J. Saikia\inst{4}
\and Martijn S.\,S.\,L. Oei\inst{5}
\and Huub J.\,A. R\"ottgering\inst{6}
}
  
%\authorrunning{abc et al.}
\institute{$^{1}$Astrophysics Division, National Centre for Nuclear Research, Pasteura 7, 02-093 Warsaw, Poland\\
$^{2}$Instituto de Astrof\' isica de Canarias, Calle V\' ia L\'actea, s/n, E-38205, La Laguna, Tenerife, Spain\\
$^{3}$Universidad de La Laguna (ULL), Departamento de Astrofisica, La Laguna,
E-38206, Tenerife, Spain\\
$^{4}$Inter-University Centre for Astronomy and Astrophysics (IUCAA), Pune 411007, India\\ 
$^{5}$Cahill Center for Astronomy and Astrophysics, California Institute of Technology, 1216 E California Blvd, CA 91125 Pasadena, USA\\
$^{6}$Leiden Observatory, Leiden University, Niels Bohrweg 2, 2300 RA Leiden, The Netherlands\\
}

% \abstract{}{}{}{}{} 
% 5 {} token are mandatory
 \date{\today} 
 
 \abstract
%   % context heading (optional)
%   % {} leave it empty if necessary  
{To test the hypothesis that megaparsec-scale giant radio galaxies (GRGs) experience multiple epochs of recurrent activity leading to their giant sizes and to understand the nature of double-double radio galaxies (DDRGs), we have built the largest sample of giant DDRGs from the LOFAR Two Metre Sky Survey (LoTSS) data release 2. This sample comprises 111 sources, including 76 newly identified DDRGs, with redshifts ranging from 0.06 to 1.6 and projected sizes between 0.7 Mpc and 3.3 Mpc. 
We conducted a detailed analysis to characterise their properties, including arm-length ratios, flux density ratios of pairs of lobes, and misalignment angles. These measurements allow us to study the symmetry parameters, which are influenced by the immediate and large-scale environments of DDRGs. Our study shows that based on the observed asymmetries of the inner lobes, the cocoons in which the inner lobes of DDRGs grow are often (approximately about 26\%) asymmetrically contaminated with surrounding material from the external medium. Our analysis also reveals highly misaligned DDRGs, which could be due to environmental factors and/or changes in the supermassive black hole jet ejection axes.  By studying the misalignment angles, we assess the stability of the jets in these systems in relation to their environment. 
For the first time, we systematically characterised the large-scale environments of DDRGs, identifying their association with dense galaxy clusters and revealing the influence of `cluster weather' on their morphologies. We have discovered a DDRG in a distant galaxy cluster at $z\sim$\,1.4. Our findings empirically confirm that dynamic cluster environments can induce significant misalignment in DDRGs, which aligns with previous simulation predictions and offers insights into how cluster weather shapes their morphology. 
Additionally, we have identified two gigahertz peaked-spectrum (GPS) candidates in the unresolved cores of the DDRGs, as well as one triple-double candidate, which, if confirmed, would be only the fifth known case.
Overall, this study enhances our understanding of the life cycle of radio AGNs and underscores the critical role of the environment in shaping the properties and evolution of giant DDRGs.}


\maketitle

\section{Introduction} \label{sec:intro}
One of the most striking pieces of evidence for restarted activity in active galactic nuclei (AGNs) can be found in jetted radio-loud AGNs or radio galaxies (RGs). This activity often manifests as a distinctive ``double-double'' radio morphology, characterised by the presence of two or more pairs of radio lobes situated on opposite sides of the radio core. Observations of these morphologies indicate that the AGN underwent a period of inactivity, followed by a reactivation after a certain hiatus. These unique sources are referred to as double-double radio galaxies (DDRGs), which were first studied by \citet{SchoenmakersDDRG1}.
Initial studies suggested that DDRGs predominantly occur in more luminous sources, such as the Fanaroff–Riley type II (FRII) radio galaxies \citep{Fanaroff1974}. However, this observation may stem from a bias in the earlier samples towards these more luminous sources, where it is easier to distinguish the two distinct pairs of lobes. Studies have shown that not all restarted sources necessarily become DDRGs, suggesting that DDRGs or restarting jets could be a normal but brief phase in the life cycle of a radio-loud AGN (RLAGN). Intermittent activity in RLAGNs can also manifest in different morphologies and at different spatial scales, as seen in recent results from \citet{Timmerman2022} for Hercules A and \citet{Kukreti2022} for 3C293. Relic radio emission from an earlier cycle of activity has also been observed in a number of compact steep-spectrum sources \citep[see][for a review]{2021OdeaSaikiaRev}. 

The spectral analysis of DDRGs reveals differences in the radio spectra of the outer and inner sources. For instance, in B1834+620, the outer lobe's spectrum is steeper than the inner lobe's, suggesting different ages or physical conditions \citep{SchoenmakersDDRG3}. The larger outer lobes often lack hotspots, implying they are no longer supplied with energy from the AGN and are older, whereas the inner lobes, being younger, exhibit characteristics of active jet supply. In RGs, the characteristics of the jets, like their stability, overall structure, and the speed at which jet-head advances, are all influenced by the properties of the surrounding medium through which the jets travel. Our current understanding of these objects comes from several studies undertaken in the past $\sim$\,20 years \citep[e.g.,][]{Saripalli2003, Saikia2006, Safouris2008, Konar2013, Konar2013b, Nandi2019}. A review on restarted radio galaxies with studies prior to 2009 can be found in \citet{Saikia2009} and updated reviews in \citet{Mahatma2023review,Morganti2024}. DDRGs provide valuable insights into the longevity and stability of jet-producing mechanisms in AGNs. The large sizes and complex structures of giant (projected total size $>$0.7 Mpc) DDRGs (G-DDRGs) suggest prolonged periods of activity interrupted by significant quiescent phases. The study of DDRGs contributes to understanding the evolution of radio galaxies, the life cycle of AGNs, and the interaction between AGNs and their environments.

The duty cycle refers to the proportion of time radio galaxies spend in different phases, including active, remnant, and restarted stages. Specifically, it allows us to examine how frequently these different phases occur and their respective duration within the life cycle of a radio galaxy. The analysis of duty cycles is crucial for understanding the evolutionary processes of radio galaxies and the impact of their jets on galactic environments \citep[e.g.,][]{Shabala2020, Turner2018}.

Numerical simulations of \citet{ClarkeBurns1991}  explore the dynamics and emission properties of restarting jets in FRII radio sources. Their simulations reveal that restarted jets are typically denser and advance faster than the original jets, though with lower Mach numbers due to the hotter, less dense cocoon material. Their study suggests that restarted jets are generally dimmer than original jets due to weaker internal shocks.

The analytical model of \citet{Kaiser2000} pointed out that the low densities in the outer lobes created by old jets would not be sufficient to explain the restarted jets' properties. They proposed that these remnant lobes interact with warmer clouds in the intergalactic medium (IGM), leading to the formation of stronger shocks and bright hotspots, thus making the restarted jets observable in radio surveys.

Simulations of \citet{Walg2014} show how episodic jet activities in AGNs, like those observed in DDRGs, involve complex, multi-phase dynamics. Each phase—from the initial jet’s progression through the undisturbed intergalactic medium to the restarted jet navigating the altered cocoon left by the initial jet—presents unique behaviours. These phases are not just sequential steps but interact dynamically, with the aftermath of one phase setting the stage for the next. The study highlights significant structural changes in AGN jets over various phases. The initial jet forms a cocoon through interaction with its surroundings. The restarted jet, moves through an already-formed cocoon, resulting in different propagation dynamics. The altered conditions in the disturbed cocoon—specifically the increased pressure and reduced density—fundamentally change the interaction between the jet and its surroundings. This impacts both the structure and propagation speed of the jet, as well as the mixing among the various gas phases within the cocoon.

As evident from the above, several models and theories have been proposed to explain the nature of DDRGs. However, these theories remain largely untested observationally due to the lack of large samples of DDRGs and the absence of focused studies using these samples. Finding and robustly analysing large samples of DDRGs is crucial, as it will not only help us understand the duty cycle of radio galaxies but also uncover a variety of complex cases with varied morphologies. Therefore, identifying large samples of DDRGs and conducting comprehensive studies to test the proposed models are of utmost importance.

The early discoveries of DDRGs were predominantly among exceptionally large sources, specifically giant radio galaxies (GRGs) with projected sizes exceeding 0.7 megaparsecs (Mpc). Notably, most known DDRGs also exhibit these giant sizes.
One hypothesis used to explain the exceptional megaparsec-scale sizes of GRGs is that they undergo multiple recurrent activity cycles, allowing them to grow larger (for a detailed review of GRGs, see \citet{GRSreview} and references therein). However, not all restarted radio galaxies necessarily exhibit the archetypal DDRG morphology, owing to differing duty cycles and AGN accretion rates. Hence, a systematic morphological analysis of GRGs is essential to quantify the fraction that display DDRG characteristics and to study their properties. This study aims to quantify the occurrence of the restarted phenomenon within the GRG population. This study, conducted under the project SAGAN\footnote{\url{https://sites.google.com/site/anantasakyatta/sagan}} \citep{D17,sagan1}, is motivated by the aim to elucidate the formation and evolution of GRGs. The results are anticipated to shed light on the broader astrophysical processes that influence the life cycle and morphology of radio AGNs, as well as the effects of their environment on these galaxies.

Throughout this paper, we adopt the flat $\Lambda$CDM cosmological model based on the Planck results, with parameters $H_0$ = 67.8 km s$^{-1}$ Mpc$^{-1}$, $\Omega_m$ = 0.308, and $\Omega_{\Lambda}$ = 0.692 \citep{2016A&A...594A..13P}. We use the convention \( S_{\nu} \propto \nu^{-\alpha} \), where \( S_{\nu} \) represents the flux density at frequency \( \nu \) and \( \alpha \) denotes the spectral index. Coordinates are in the J2000 coordinate system.

\section{Methodology for identifying giant DDRGs}\label{sec:sample}
Lobes of radio galaxies become fainter over a period of time due to adiabatic,  inverse Compton and synchrotron energy losses of the electrons. Hence, identifying DDRGs can be challenging as it requires the radio surveys/images to be sensitive to diffuse emission and have good spatial resolution. Only under these conditions can the two distinct episodes and their respective extents be unequivocally pinpointed and effectively categorised \citep[for more, see;][]{Mahatma2023review}. 

Prior to the advent of very sensitive low-frequency surveys like the LOFAR Two Metre Sky Survey \citep[LoTSS;][]{lotsspdr,lotssdr1} at 144 MHz, very few DDRGs were known or identified with diffuse outer lobes. LoTSS's exceptional sensitivity to low--surface brightness diffuse emission and adequate spatial resolution render it highly suitable for identifying and studying DDRGs. The studies of \citet{Mahatma2019} and \citet{PDLOTSS} using LoTSS DR1 data (image resolutions of 6\arcsec and 20\arcsec) found about 40 DDRGs with low flux densities and diffuse emission. Deeper surveys with LOFAR of the Lockman Hole field also unveiled low--surface brightness DDRGs \citep{Jurlin2020}, which would have been difficult to observe or identify in higher-frequency surveys. The low-frequency nature of this survey, combined with its high sensitivity, enables the detection of a lot of bridge emissions\footnote{Usually referred to the diffuse faint emission between radio core and lobes, for more see \citet{Leahy1984}.} in radio galaxies.

In our study, we sought to determine the proportion of GRGs that exhibit indications of restarted activity, specifically demonstrating DDRG morphology. LoTSS DR2 \citep{Shimwell2022} with its higher sensitivity (median rms sensitivity of 83 \mujybeam~ for 6\arcsec~ resolution maps) is more prone towards the detection of diffuse and aged emissions that are typically elusive at higher frequencies.

\begin{figure*}
  \includegraphics[scale=0.3]{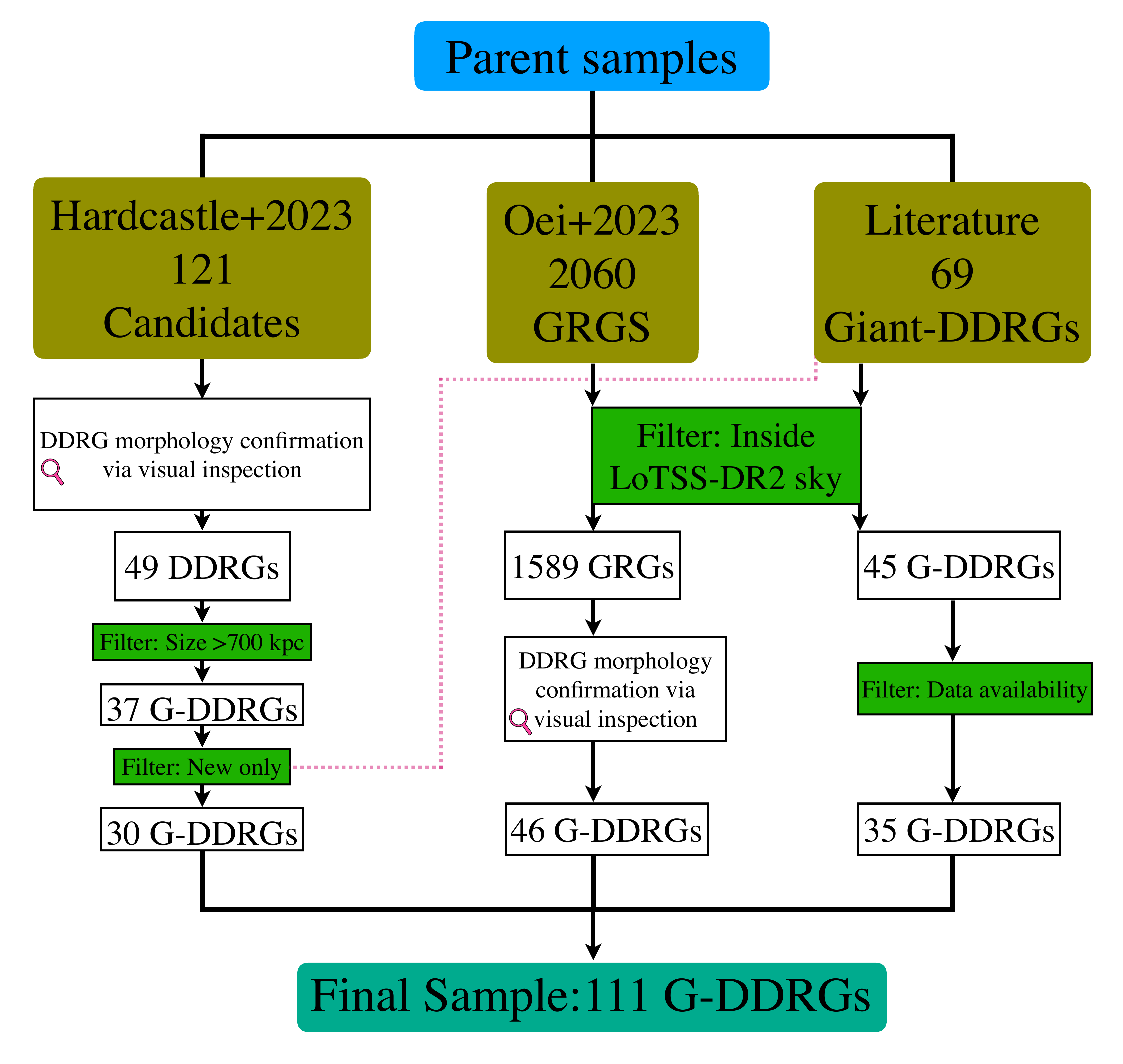}
  \caption{\label{fig:flowchart} Flowchart illustrating the process for creating the G-DDRG catalogue utilised in our analysis. Here, Hardcastle+2023 refers to the catalogue from \citet{Hardcastle2023}, and Oei+2023 indicates the catalogue from \citet{oei2023}. 69 G-DDRGs from the `Literature' category are previously reported DDRGs (see reference in Tab.~\ref{tab:main}).}
\end{figure*}

The identification was carried out using candidate sources from three parent samples within the LoTSS-DR2 sky coverage. The workflow process is illustrated in Fig.~\ref{fig:flowchart}. These parent samples are derived from the works of \citet{oei2023}, \citet{Hardcastle2023}, and a compilation of confirmed DDRGs from existing literature. For the third parent sample, which already consists of confirmed DDRGs, the objective was to determine how many are located within the LoTSS-DR2 sky area and show DDRG morphologies (i.e., two pairs of radio lobes) at the angular resolution of the survey. This comprehensive approach allows for a thorough and efficient search, enhancing our ability to find as many DDRGs as possible. 

To carry out the morphological examination of the sources, we utilised radio maps from LoTSS DR2, the Faint Images of the Radio Sky at Twenty-Centimeters (FIRST) survey \citep{FIRST1995Becker}, and the Very Large Array Sky Survey \citep[VLASS;][]{VLASS}. The VLASS offers an image resolution of $\sim$2.5\arcsec ~at 3~GHz, facilitating the resolution or detection of jet/lobe structures (inner) of approximately kiloparsec (kpc) scale in host galaxies with redshifts of $\sim$\,0.02. Radio surveys, FIRST (1.4~GHz) and VLASS (3~GHz), with their higher frequencies and resolutions, serve as complementary tools to LoTSS in identifying radio cores, especially in cases where LoTSS may not have detected them. We have only considered sources where inner lobes are clearly distinguishable and optical identification is possible.

Firstly,  we meticulously inspected the radio maps of all GRGs listed in the catalogue by \citet{oei2023}. This catalogue encompasses 1589\footnote{The original catalogue reports 2,060 sources, but 471 of these lie in a sky area outside the DR2 region, which is not publicly available at present.} GRGs identified from the LOFAR Two-meter Sky Survey (LoTSS) Data Release 2 (DR2), \citep{Shimwell2022}, which spans an area of $\sim$\,5600 \sqdeg ~at 144~MHz. This led to the identification of 46 GRGs that showed morphological characteristics of DDRGs or G-DDRGs. 

Secondly, we utilized the catalogue by \citet{Hardcastle2023} from LoTSS DR2, which presented 121 candidate DDRGs. Through rigorous manual verification of these candidates, using all available radio and optical maps, we reliably identified 30 new G-DDRGs.

Finally, there are about 69 GRGs known with DDRG morphology, which form our literature parent sample as shown in Fig.~\ref{fig:flowchart}  \citep{Schoenmakers1999GPS, SchoenmakersDDRG3, SchoenmakersDDRG1, Schilizzi2001, Saripalli2002, Saripalli2003, Marecki2003, Machalski2006, Saikia2006, Nandi2012, Saripalli2013, Renteria2017, Kuzmicz2017, Mahatma2019, PDLOTSS, sagan1, Kozielrogue2020,  Simonte2022}.
Of the 69 GRGs, 45 fall within the LoTSS DR2 survey area. For uniformity in our sample selection and analysis, we only included sources exhibiting DDRG morphologies in the radio maps from LoTSS DR2. Specifically, we excluded DDRGs identified by \citet{Simonte2022} from deeper LOFAR surveys and those classified using very high-resolution VLBI data from the literature. This approach ensures consistency in our data, resulting in a final sample of 35 G-DDRGs from the literature.

For robust verification and analysis (see Sec.~\ref{sec:analysis}) of all candidate DDRGs, including re-verification of known DDRGs from the literature, we used comprehensive radio data (surveys mentioned above) and optical images from the Sloan Digital Sky Survey (SDSS; \citealt{sdss00, sdssdr14}), the Panoramic Survey Telescope and Rapid Response System (Pan-STARRS; \citealt{chambers16}), and the DESI Legacy Imaging Surveys \citep{Dey2019Legacy}. Spectroscopic redshift measurements were obtained from SDSS DR16. For sources lacking spectroscopic redshifts, photometric redshifts were taken from \citet{Duncan2022}. This approach ensured the reliability and accuracy of our final G-DDRG sample. 

\begin{figure*}
  \includegraphics[scale=0.29]{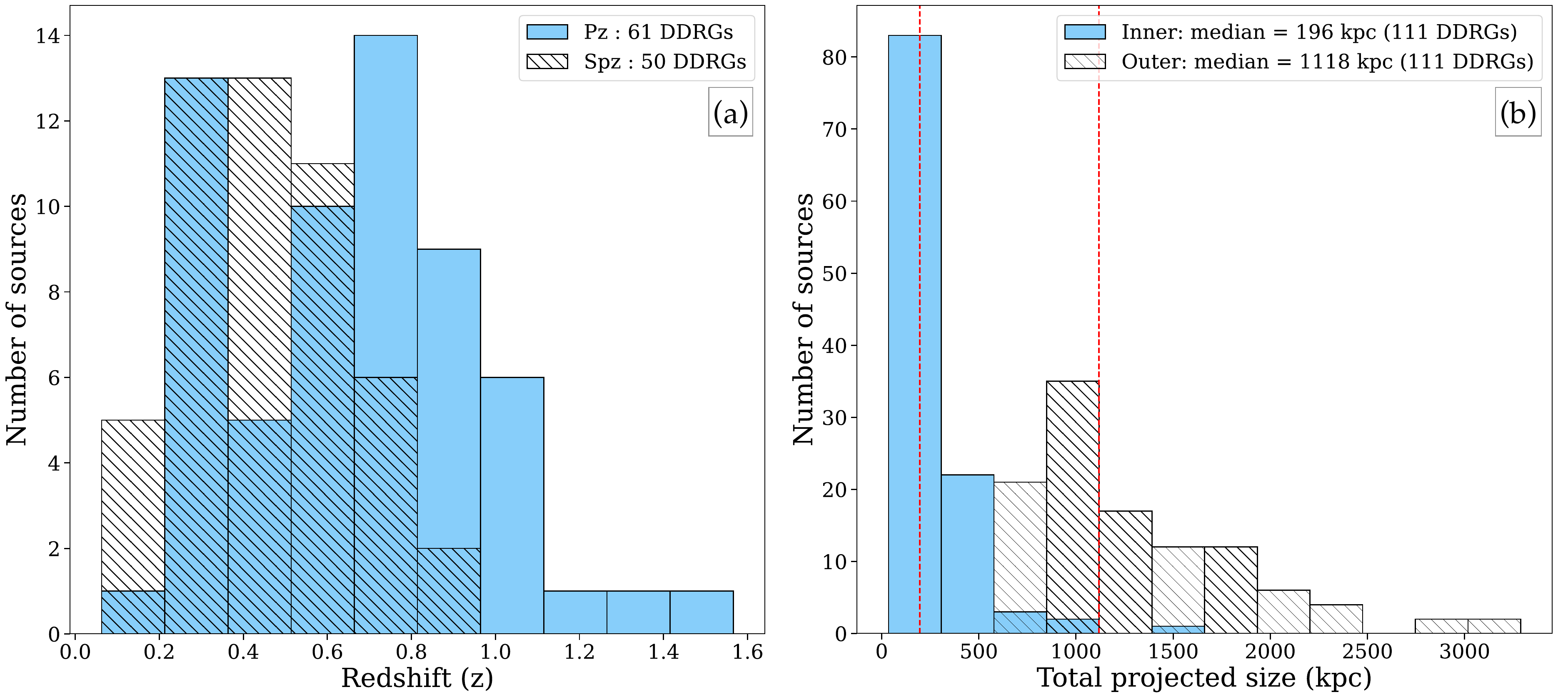}
  \caption{\label{fig:zsizedist} 
 \textit{Left}: Distributions of the redshift of the sample of G-DDRGs with spectroscopic (Spz) and photometric (Pz) measurements shown with hatched and unhatched bins, respectively. \textit{Right}: Distributions of projected sizes, where the red dashed lines indicate the median values.}
\end{figure*}

Consequently, this resulted in the compilation of a final catalogue comprising 111 confirmed G-DDRGs from the LoTSS DR2 sky area. 
The distribution of their redshifts and projected sizes are shown in Fig.~\ref{fig:zsizedist}. The redshift range of sources in our sample is \( 0.06289 \leq z \leq 1.566 \), with 50 spectroscopic measurements and the rest photometric. The sizes of the inner doubles range from 35 kpc to 1.5 Mpc, while the outer doubles range from 0.7 Mpc to 3.3 Mpc.
All the sources with their properties are presented in Tables ~\ref{tab:main} and ~\ref{tab:2}. Montages of their radio images are presented in figures~\ref{fig:Mosaic1} to ~\ref{fig:Mosaic4}.

\begin{figure*}
  \includegraphics[scale=0.3]{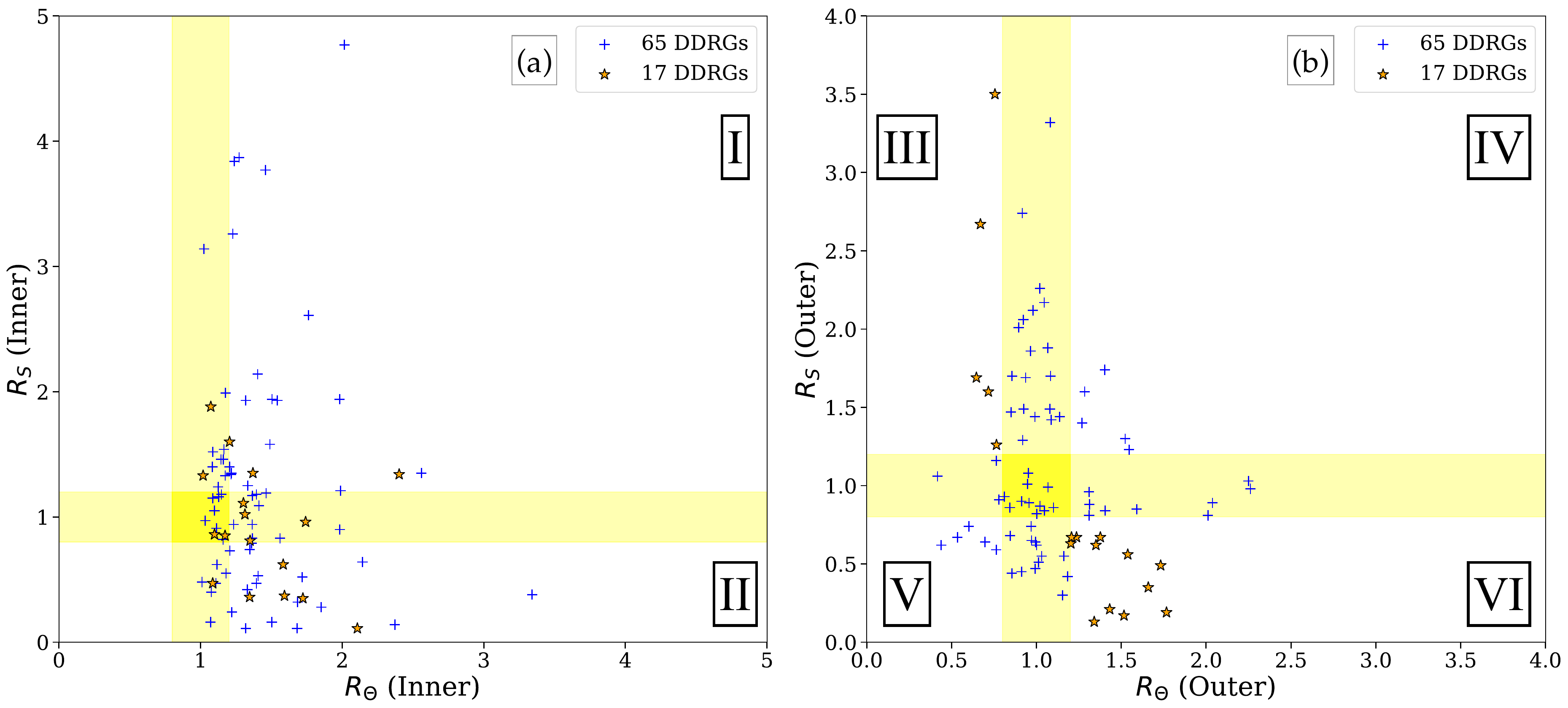}
  \caption{\label{fig:FALRATIO} $\rm R_{s}$ v/s R$_{\theta}$ (flux density ratio v/s arm-length ratio) plot for inner doubles (left plot) and outer doubles (right plot). Star markers in orange represent sources for which the shorter of the two outer lobes is brighter, indicating the interaction of outer lobes with the environment. The remaining DDRGs are shown with the blue `+' marker. Sources with their measurements falling in the yellow bands of both figures are considered to be symmetric (flux or arm-length). For more details see Sec.~\ref{sec:sympara}.}
\end{figure*}

\begin{figure}
  \includegraphics[scale=0.47]{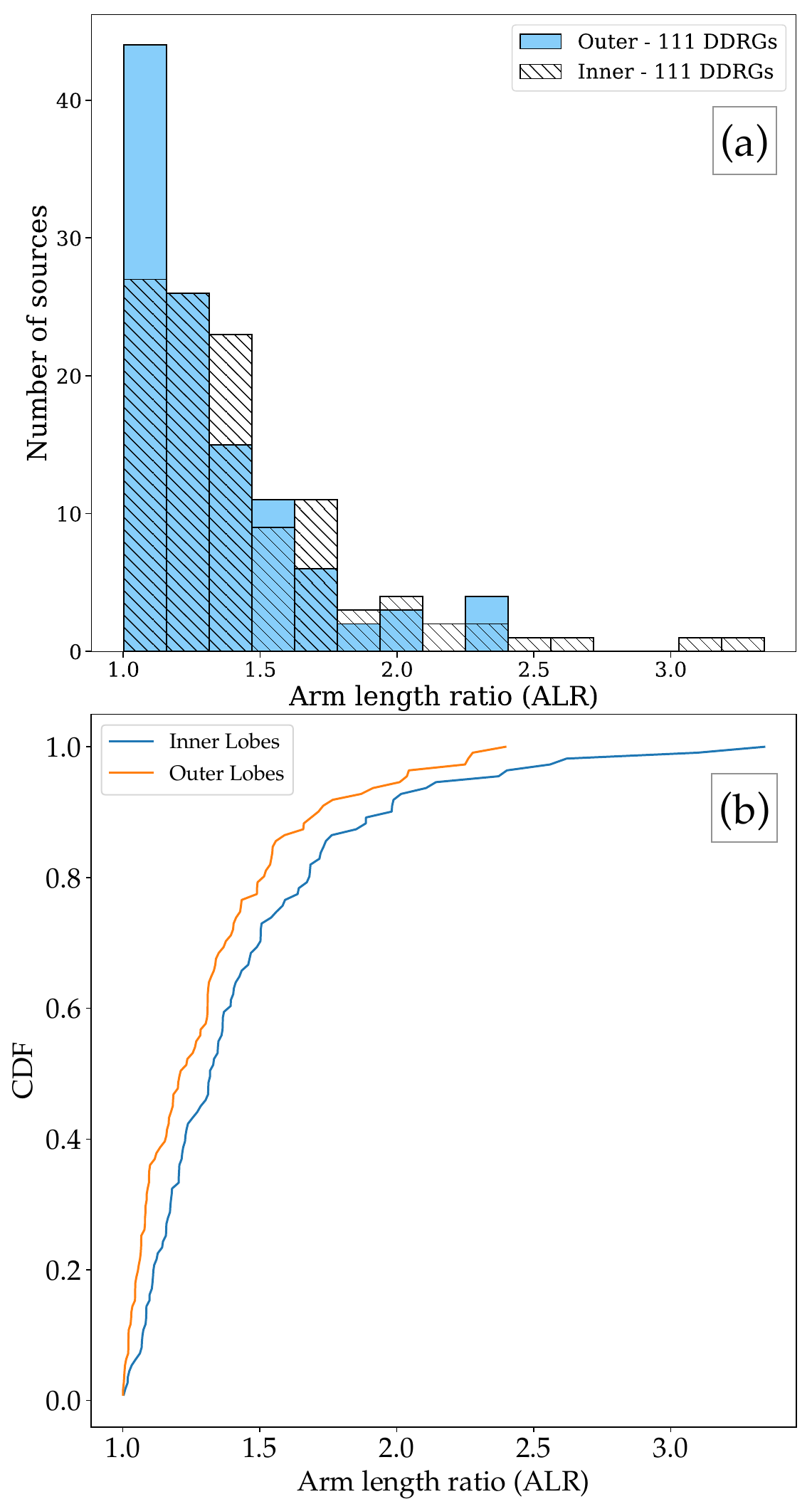}
  \caption{\label{fig:alrcorr} (a) The Upper figure shows the histogram distributions of arm-length ratios (ALR) of inner and outer lobes, where ALR is defined as greater than 1. (b) The lower figure shows the cumulative distribution function (CDF) of arm-length ratios for the inner (blue) and outer (orange) lobes. For more details see Sec.\ref{sec:sympara}.}
\end{figure}

\begin{figure*}
  \includegraphics[scale=0.3]{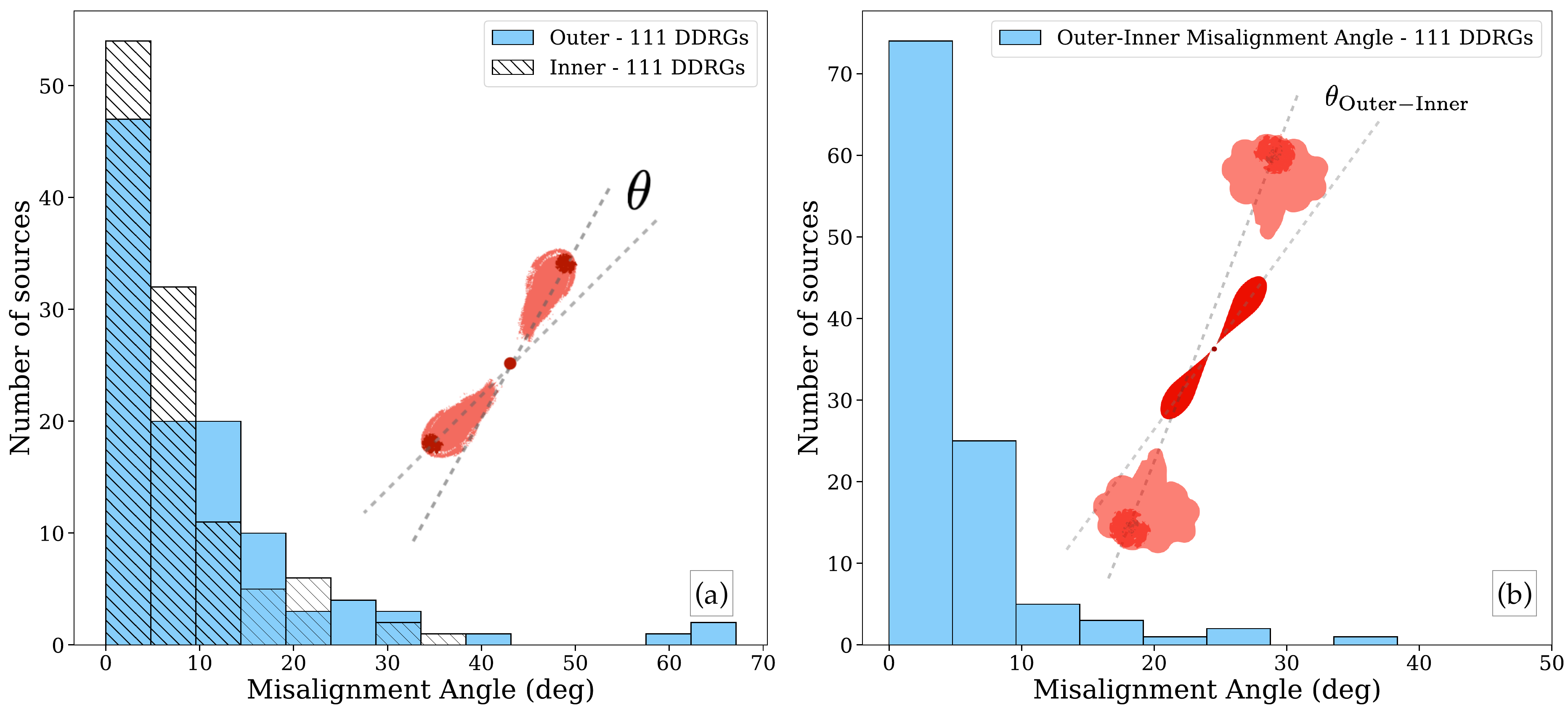}
  \caption{\label{fig:MA} \textit{Left:} Distribution of misalignment angle between respective lobes (inner and outer) on each side of the core, where median Outer misalignment = 6.9\deg and median Inner misalignment = 4.8\deg. \textit{Right:} Distribution of misalignment angle between outer and inner lobes (median=2.9\deg). The diagrams in each plot illustrate one of many possible cases, serving as a general representation of how measurements for misalignment angles were taken (for more details, see Sec.~\ref{sec:MA}).}

\end{figure*}

\section{Analysis}\label{sec:analysis}
Given the complicated morphologies of these sources, careful manual inspection and measurements are necessary to avoid potential errors. This meticulous process has been carried out for a large sample, enabling us to derive significant inferences about the nature of these sources.
The following steps are followed for systematic and consistent measurements of the sizes of the inner and outer lobes.

\begin{itemize}
    \item Inner lobes: Each typically exhibits a single distinct peak, which is used to estimate the corresponding angular size.

    \item Outer lobes (a): If the outer lobes have a single prominent peak (for example, as observed in the radio image of J084525+522915), we note the positions of these peaks to determine the corresponding angular size. 
    
    \item Outer lobes (b): For outer lobes that appear diffuse as for example in J150301+572318, we identify the location farthest from the core with a brightness value exceeding 9$\sigma$, where $\sigma$~ represents the root mean square (rms) noise of the image. We mark one such location in each outer lobe to ascertain the angular size.

    \item Outer lobes (c): In scenarios where the outer lobes are so diffuse as for example in J235751+332608, that no brightness value surpasses the 9$\sigma$ threshold, we select the local brightness maximum above 3$\sigma$.

\end{itemize}
The number of lobes in the above three categories are 143, 47 and 32 respectively.
The corresponding angular and projected sizes can be found in Columns 6 and 7 of Tab.~\ref{tab:main}.

Further, we have measured the flux densities of the inner and outer lobes, as well as of the core and the core spectral index. The flux densities were estimated using Common Astronomy Software Applications \citep{CASA} with the task {\tt CASA-VIEWER} by manually selecting the appropriate regions. Additionally, we have calculated the arm-length ratios, flux density ratios of the lobes, and misalignment angles. Further details are provided in the subsequent sections.

\section{Results}
In this section, we present our findings related to the symmetry parameters of the giant DDRG sample using the arm-length and flux density ratios of the inner and outer lobes. We present our diagnostic diagram designed to best utilise these properties to study the symmetry parameters of these sources.

\subsection{Arm-length and flux density ratios} 
Arm-length ratio (ALR or $\rm R_{\theta}$) is an important property for understanding the physical characteristics and dynamics of radio galaxies. The ALR for the inner lobes is calculated by dividing the angular size of the longer arm by the angular size of the shorter arm. The same sense of direction was used to measure ALR for the outer lobes. Their values for our sample are presented in Columns 11 and 12 of Tab.~\ref{tab:main}.
These ratios are crucial in assessing the symmetry of radio galaxies and understanding their evolution. They can indicate asymmetries in the jet outflow, environmental asymmetries or orientation effects. A near-unity ALR suggests a more symmetric jet ejection and propagation environment and a large inclination angle.
Similar to ALR or $\rm R_{\theta}$, the flux density ratios ($\rm R_{s}$) of the inner and outer doubles were estimated which are listed in columns 3 and 4 of Tab.~\ref{tab:2}, respectively. For $\rm R_{s}$ measurements, we excluded sources where the inner doubles were not well-resolved or partially embedded in the outer lobes. Thus, these properties were only analysed for sources with well-separated inner and outer doubles. Consequently, we were able to reliably determine the flux density ratios for the inner and outer lobes in 82 sources.

\subsection{Symmetry Parameters of giant DDRGs}\label{sec:sympara}
In a typical radio AGN, bipolar jets \citep[see][for a recent review]{Saikia2022} are likely to be emitted with equal thrust in both directions, suggesting that their propagation through the external medium should lead to symmetric structures unless affected by environmental factors. If one lobe or jet-head propagates through a denser medium than the other, it will have a shorter distance to the radio core (shorter arm) and may appear brighter. Observing a lobe with a shorter arm that is also brighter provides evidence of environmental asymmetry on opposite sides of the radio core.
Comparing the symmetry parameters of the inner and outer doubles can yield information on the conditions through which the jets traverse and highlight any inherent asymmetries in the jet structures. For restarted jets advancing into cocoons from previous activity epochs, one would anticipate encountering similar conditions, resulting in more symmetric structures for the inner doubles. However, \citet{Saikia2006}, using a sample of 12 DDRGs, found that the inner doubles exhibit greater asymmetry in terms of arm-length and flux density ratios (R$_{\rm \theta}$ \& $\rm R_{s}$)  compared to the outer doubles. The asymmetry could be attributed to variations in environmental conditions, degrees of entrainment in the cocoons, or intrinsic jet asymmetries. Additionally, if the inner doubles appear more collinear than the outer doubles, it possibly reflects the large-scale density gradients affecting the outer doubles.
In our current study, we have the opportunity to examine these findings with a larger sample to identify any general trends associated with DDRGs. Specifically, we can investigate questions such as: Does the initial jet activity create a less dense environment that facilitates the propagation of subsequent jets? If so, would this result in consistently higher symmetry in the newer jet structures? Our larger sample allows us to investigate these possibilities in greater detail and determine if these trends hold true across a broader range of DDRGs.

We introduce a diagnostic scatter plot (Fig.~\ref{fig:FALRATIO}) of $\rm R_{\theta}$ and $\rm R_{s}$ for inner and outer lobes to examine the symmetry parameters of the DDRGs. 
$\rm R_{\theta}$ close to unity indicates less environmental effects or more stability in jet propagation, whereas significant deviations in the outer $\rm R_{\theta}$, especially when not mirrored in the inner $\rm R_{\theta}$, could suggest that factors like environmental asymmetries or orientation effects are at play. 
The diagnostic plot aims to assess whether the asymmetry in the environment was mitigated by the first epoch doubles. Specifically, if the outer lobe asymmetry (R$\rm _{\theta(outer)}$) is present but the inner lobe (R$\rm _{\theta(inner)}$) is symmetric, it suggests that the outer doubles may have cleared the asymmetric environment. Conversely, if the inner doubles exhibit asymmetry while the outer doubles do not, it implies contamination of the cocoons by the environment.
The results from the plots can be summarised as follows:

\begin{itemize}
    \item R$\rm _{\theta(inner)}$ is always greater than 1 for all sources because this ratio is calculated by dividing the length of the longer arm by the length of the shorter arm (Regions I and II). To maintain consistency, we have used the same directional symmetry for calculating R$\rm _{\theta(outer)}$, regardless of which arm is longer or shorter (see Fig.~\ref{fig:FALRATIO}b). 

    \item The yellow regions in the plot represent sources with $\rm R_{s}$  and $\rm R_{\theta}$ values between 0.8 and 1.2, to help distinguish the significantly asymmetric sources.
  
    The G-DDRGs are all associated with galaxies and in the framework of the unification scheme are likely to be inclined at large angles to the line of sight. \cite{Barthel1989ApJ...336..606B} suggested that radio galaxies are inclined at angles larger than about 45$^\circ$, while quasars are at smaller angles. For a velocity of advancement of $\sim$0.1c  \citep{Scheuer1995}, the expected values of ALR or R$_\theta$ are less than about 1.1 for angles of inclination greater than 60$^\circ$. 

    \item R$_{s} >$ 1: Indicates the ratio of the brighter lobe to the fainter lobe. R$_{s} <$ 1: Indicates the ratio of the fainter lobe to the brighter lobe.

    \item Sources with asymmetry in their outer lobes where the nearer outer lobe is brighter are highlighted with star markers in Fig.~\ref{fig:FALRATIO}b, and these same sources are mapped in Fig.~\ref{fig:FALRATIO}a. The use of star markers helps emphasise sources that experience environmental asymmetry in their outer lobes and compare them with the newer or inner epoch properties.

     \item Star-marked sources on the Fig.~\ref{fig:FALRATIO}a, that fall within yellow bands, indicate that in these instances, the environmental asymmetry seems to have been mitigated by the action of the outer lobes or during the first episode of jet activity. The ability of outer doubles to clear asymmetric environments is crucial for understanding the evolution of DDRGs. This clearing effect can mitigate asymmetry in subsequent jet activity, leading to more symmetric inner doubles.
\end{itemize}

We mark key regions of the two plots (a and b of Fig.~\ref{fig:FALRATIO}) with six regions which are described below.

\begin{itemize}

    \item Region {I}: The inner lobe farther away (longer arm) from the core is brighter.
    \item Region II: The inner lobe closer (shorter arm) to the core is brighter ($\sim$\,26\%). These sources highlight a clear case of environmental asymmetry impacting the inner doubles. 
    \item Region III: The outer lobe closer to the core is brighter, possibly indicating the jet environment interaction. Shows sources with flipped symmetry for the inner and outer pairs lobes. These sources are marked by stars.
    
    \item Region IV \& V: The outer lobe farther away from the core (longer arm) is brighter. 
    \item Region VI: Contains sources marked by stars where the shorter arm is brighter (asymmetric environment).
    \item Stars in region II suggest that the same environmental asymmetry affecting the outer lobes also impacts the inner lobes. 

\end{itemize}

Greater asymmetry in inner doubles compared to outer doubles could result from an intrinsic asymmetry in the cocoon on opposite sides of the nucleus. Newer inner lobes are more affected by current environmental conditions, leading to increased asymmetry. This suggests that the inner doubles reflect recent changes in the surrounding medium, while the asymmetry in the outer doubles reflects conditions in the large-scale environment.

The study by \citet{Kaiser2000} proposed that material from the surrounding IGM penetrates the cocoon boundary. This material interacts with the jets. Their analytical models indicate that the gas densities within the old cocoon are insufficient to slow down the jet, necessitating the influx of external material. This entrained IGM material could modify the densities, thereby slowing the jets and forming hotspots. 
As mentioned earlier in this section, if the cocoon in which the newer or inner lobes are growing is asymmetrically contaminated, it could lead to structural asymmetries, which we observe in several of our sources. 
 
The contamination of the cocoon, as suggested by \citet{Kaiser2000}, can arise from several factors: (1) Kelvin-Helmholtz instabilities that can arise from velocity shear between the jet and the cocoon material.
(2) Rayleigh-Taylor instabilities can occur at the interface of the cocoon and the external medium due to differences in density. (3) Buoyancy Effects: These effects can contribute to the mixing of cocoon material with the surrounding medium. (4) Shredding of Warm Clouds: This mechanism involves warm ($\sim 10^{4}$~K) clouds of gas being shredded by the bow shock of the jet, leading to the penetration of these clouds into the cocoon. This can be considered a likely mechanism for cocoon contamination, as proposed by \citet{Kaiser2000}. This process can explain the presence of sufficient gas density within the cocoon to support the formation of inner lobes and prevent quasi-ballistic propagation of the new jets.

In instances where the lobe interacts with a denser medium, there is a significant enhancement in the dissipation of energy, leading to a more efficient conversion of beam energy into radio emissions \citep[e.g.,][]{EilekShore1989,GKW1991,Blundell1999,Jeyakumar2005}. Lobes encountering a denser medium can be verified using polarisation measurements, as these would show increased depolarisation. For instance, the highly asymmetric source 3C459 exhibits significant depolarisation, indicating interaction with a denser medium \citep{Thomasson2003}. Furthermore, \citet{Arshakian2000} observed that asymmetry is more common in luminous sources.
Consistent with this understanding, our observations reveal that in several cases within our sample, the lobe proximal to the radio core, or the `shorter' lobe, exhibits greater brightness. This observation lends credence to the hypothesis that the reduced length of the lobe, and its increased brightness, are indicative of encountering resistance in its propagation, affirming its interaction with a denser medium. 

In our sample, certain sources exhibit a notable symmetry in their outer lobes, while, intriguingly, their corresponding inner lobes display asymmetry. A plausible explanation for this phenomenon could be the impact of galaxy mergers. Such mergers can result in denser ISM within the nuclear regions (the central few kpcs), potentially triggering the formation of radio jets. 
As these jets traverse through an asymmetrically/unevenly distributed gaseous environment, it is conceivable that one of the bipolar jets encounters a denser ISM compared to its counterpart. This disparity in the density of the ISM encountered by each jet could ultimately lead to the formation of asymmetric inner lobes.

Studies of Compact steep-spectrum sources (CSS) by \citet{Saikia2001,Jeyakumar2005} 
and of 3CRR sources by \citet{Arshakian2000} have shown that intrinsic or environmental asymmetries in the radio sources are more pronounced at smaller physical scales. This means that the differences in structure and environment around the radio sources are more apparent when observed at smaller physical scales.
Conversely, as the physical scale increases, the level of these asymmetries diminishes as they may be governed by different or additional factors than those at smaller scales (e.g., IGM). Similarly, our ALR analysis of the inner and outer lobes is consistent with the above as we observe the inner lobes to be more asymmetric than the outer ones. Defining the ALR or R$_\theta$ to be $>$1 for both the inner and outer lobes, the median values are 1.32 and 1.21, respectively. Their distributions can be seen in the histogram shown in Fig.~\ref{fig:alrcorr}a and the cumulative distribution function in Fig.~\ref{fig:alrcorr}b, which show the clear differences between the two distributions. Furthermore, a Kolmogorov-Smirnov test confirms the distinction between the two distributions with a p-value of 0.017. This trend not only aligns with the findings reported by \cite{Saikia2006}, but also extends the understanding of these differences through analysis of a much larger sample size, thereby providing more robust evidence.

\subsection{Misalignment angle} \label{sec:MA}
\citet{SchoenmakersDDRG1} found that the inner and outer lobes are typically aligned within $\sim$\,10\deg, suggesting a stable direction for the jet outflow over the activity periods. However, this conclusion was based on a small sample of just seven sources. With a larger sample of 111 sources, we now extend this analysis to quantify the average misalignment of inner and outer doubles individually, as well as the overall misalignment of the two pairs of lobes. Such an investigation will provide deeper insights into the influence of environmental factors on the jet morphology and the overall stability of jet directions in DDRGs.

The outer misalignment angle is defined as the angle formed by lines connecting the position of the host galaxy/radio core with the locations of the outer lobes, where the outer lobes' positions are determined based on predefined criteria given in Sec.~\ref{sec:analysis}. Likewise, the inner misalignment angle refers to the angle between lines that link the host galaxy's location with that of the inner lobes. Additionally, the inner-outer misalignment angle is characterised as the angle between two lines, each joining the lobe positions corresponding to different episodes. These two lines may not intersect at the nucleus of the galaxy due to the misalignment of each of the two pairs of lobes. The values of the misalignment angles are presented in Columns 8, 9, and 10 of Tab.~\ref{tab:main}.

We present the distributions of misalignment angles of inner, outer and inner-outer in Fig.~\ref{fig:MA}. We observe that the distributions of misalignment angles for both inner and outer are skewed towards lower angles, with a significant concentration of values below 20\deg. The median misalignment angles are 6.9\deg~ for the outer and 4.8\deg~ for the inner lobes. Considering only those with well-defined identifiable peaks in the outer lobes, the median misalignment angles for the outer is 4.8\deg. This suggests a relatively stable jet direction over the active periods. The presence of higher misalignment angles in some sources indicates that environmental factors or internal dynamics can occasionally cause significant deviations in the jet direction. 
By comparing the misalignment angle between the outer and inner doubles, one could infer whether the environment has changed significantly over time or if the jet's orientations have altered. Precession of the supermassive black hole or changes in the jet ejection axis due to interactions between the two episodes can also cause significant misalignment between the outer and inner doubles. 

Precessing jets exhibit a wide variety of structures depending on the inclination angle, cone angle and jet speed \citep[e.g.][]{Gower1982ApJ...262..478G,Horton2020MNRAS.499.5765H,Nolting2023ApJ...948...25N}. In addition to the effects of precession, interactions with the external environment may push jets away from their ballistic paths adding to the complexity of structures observed in such sources. This can lead to the line joining the prominent peaks on opposite sides not passing through the radio core. Examples of such sources where the overall structure has been suggested to be due to the precession of the jets include J0643+1044 \citep{SethiS2024} and PKS~2300$-$18 \citep{Misra2025MNRAS.536.2025M}. Only nine sources in our sample exhibit a positional configuration where the lines connecting both pairs of lobes pass directly through the core radio component or host galaxy nucleus (e.g., ILTJ105742.51+510558.5).

The analysis of standard deviations provides valuable insights into the variation in misalignment angles in DDRGs. The misalignment angle for the outer doubles exhibits the highest variation, with a standard deviation of $\sim$\,12\deg, but it reduces to $\sim$\,7\deg while considering only those with well-defined peaks. The greater misalignment angles observed for diffuse outer lobes could result from a combination of measurement uncertainties and partly due to encountering turbulent or inhomogeneous conditions for those in clusters of galaxies. The inner misalignment angles have a standard deviation of $\sim$\,7\deg, suggesting that the inner lobes are also aligned but still subject to some variations due to interactions closer to the AGN core. The outer-inner misalignment angles have a standard deviation of $\sim$\,6\deg (median=2.9\deg) for the entire sample, and $\sim$\,7\deg (median=2.0\deg) for those with well-defined outer peaks. Thus the inner and outer doubles are usually along a similar direction, and large changes are rare (Fig.~\ref{fig:MA}).

\begin{figure}
  \includegraphics[scale=0.33]{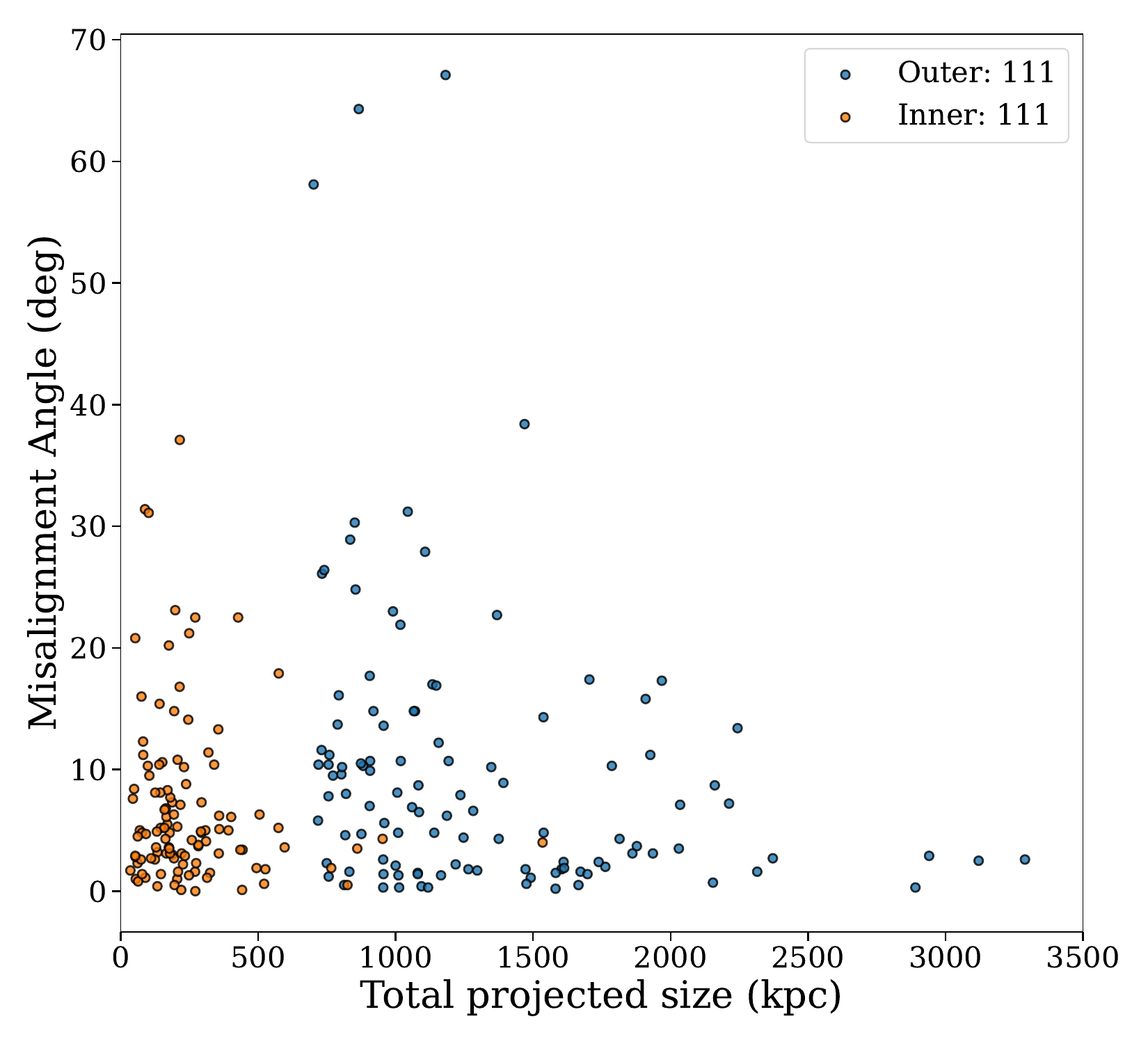}
  \caption{\label{fig:MA_vs_size} This figure shows the misalignment angle of outer and inner lobes v/s total size of the source. Outer lobes are marked by blue circles, while inner lobes are marked by orange circles.}
\end{figure}

In Fig.~\ref{fig:MA_vs_size}, we present the misalignment angles of the inner and outer doubles in relation to their projected linear sizes. The Figure shows that sources with powerful jets that reach larger ($>$1.5~Mpc) distances tend to 
maintain a stable direction and possibly reside in uniform environments. The outer doubles of these larger sources have a median misalignment angle of 2.9\deg with a standard deviation of 5.2\deg. In contrast, the outer doubles of sources with projected sizes $<$ 1.5 Mpc exhibit a higher median misalignment angle of 8.8\deg ~with a standard deviation of 13.5\deg.

\subsubsection{Individual notes on sources with extreme misalignment} \label{sec:indi}
The study of misaligned DDRGs helps in understanding the dynamics of RLAGN, providing insights into phenomena such as jet precession or orientation changes, potentially caused by binary black hole interactions \citep[e.g.,][]{Begelman1980,Campanelli2007,Saripalli2013,Nandi2017,GK_dDDRG_2022}. These processes can inform us about the evolution of galaxies and the role of AGNs in shaping them. 

DDRG J100956.00+365022.6 (ILTJ100956.68+365022.4): The misalignment angle between the inner and outer doubles of this source is $\sim$\,37\deg, the largest observed in our sample, indicating a significant offset between the axes of the inner and outer doubles. The outer lobes are notably diffuse, characteristic of relic lobes. This configuration bears a striking resemblance to the small-sized DDRG (J102228.41+500619.8) in Abell~980, also referred to as a detached DDRG \citep[see;][]{GK_dDDRG_2022}. In their study, this feature is attributed to the outer double buoyantly rising away from the centre of the intra-cluster-medium (ICM), with the DDRG hosted by the BCG of the Abell 980 galaxy cluster.

\begin{figure}
  \includegraphics[scale=0.15]{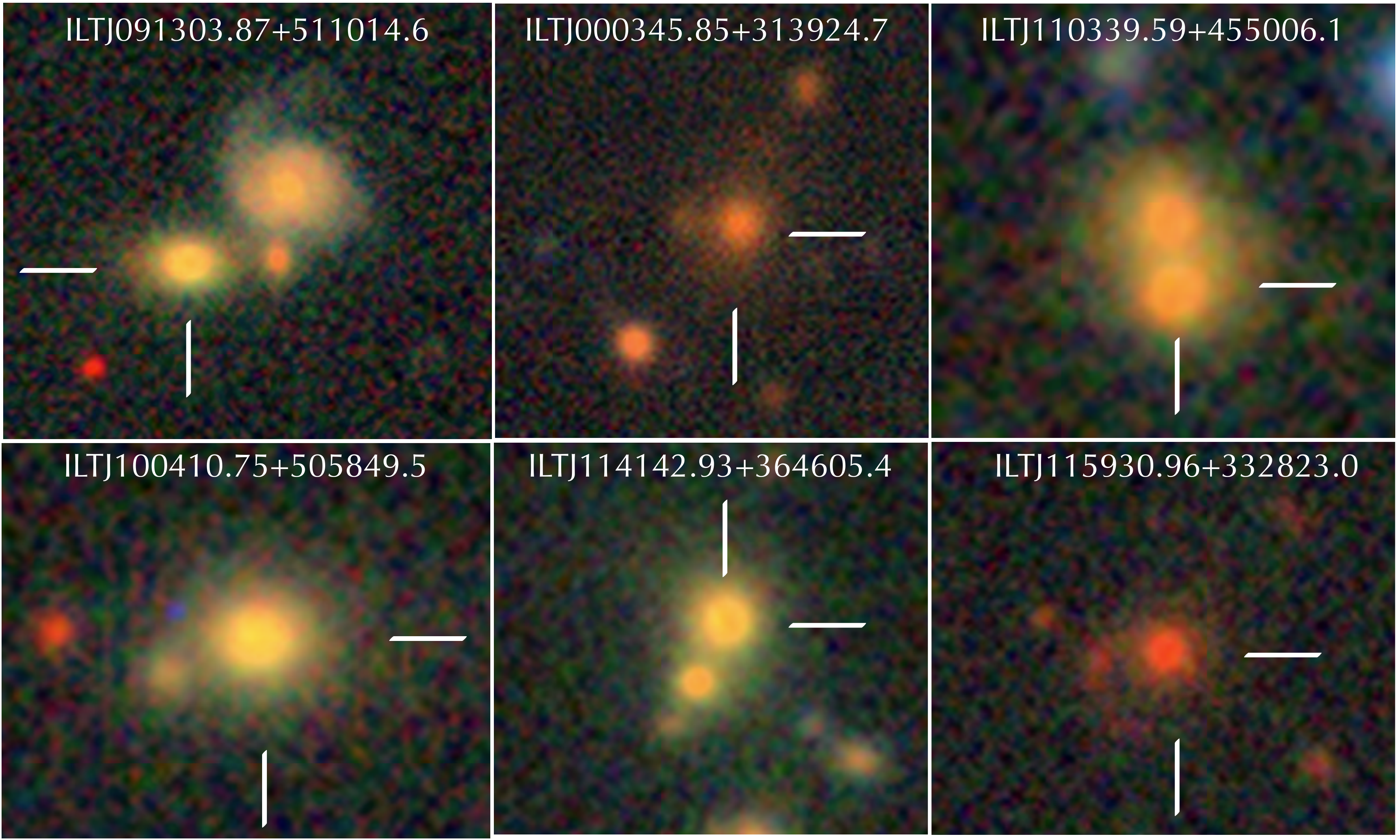}
  \caption{\label{fig:merger} The figure shows a montage of host galaxies of six  G-DDRGs which show possible signs of mergers as described in Sec.\ \ref{sec:merger}. The optical colour images are from  DESI Legacy Imaging Surveys DR9. }
\end{figure}

DDRG J091305.16+511026.23 (ILTJ091303.87+511014.6): The inner lobes are misaligned by 22.5\deg, and the outer lobes by 17.4\deg. The observed misalignment in these radio morphological features can possibly be linked to or explained by an ongoing merger in the host galaxy, as seen in the top left of Fig.~\ref{fig:merger}.

The three outliers with the misalignment angle of the outer lobes being $>$50\deg, ILTJ083201.90+395548.2, ILTJ131458.85+382709.1 and ILTJ163415.49+492846.4, have a wide-angle tailed structure and are located in clusters of galaxies (see Sec.~\ref{sec:env} and Tab.~\ref{tab:clusters}), underlining the importance of environmental effects in some sources. 

\begin{figure*}
\centering
  \includegraphics[scale=0.28]{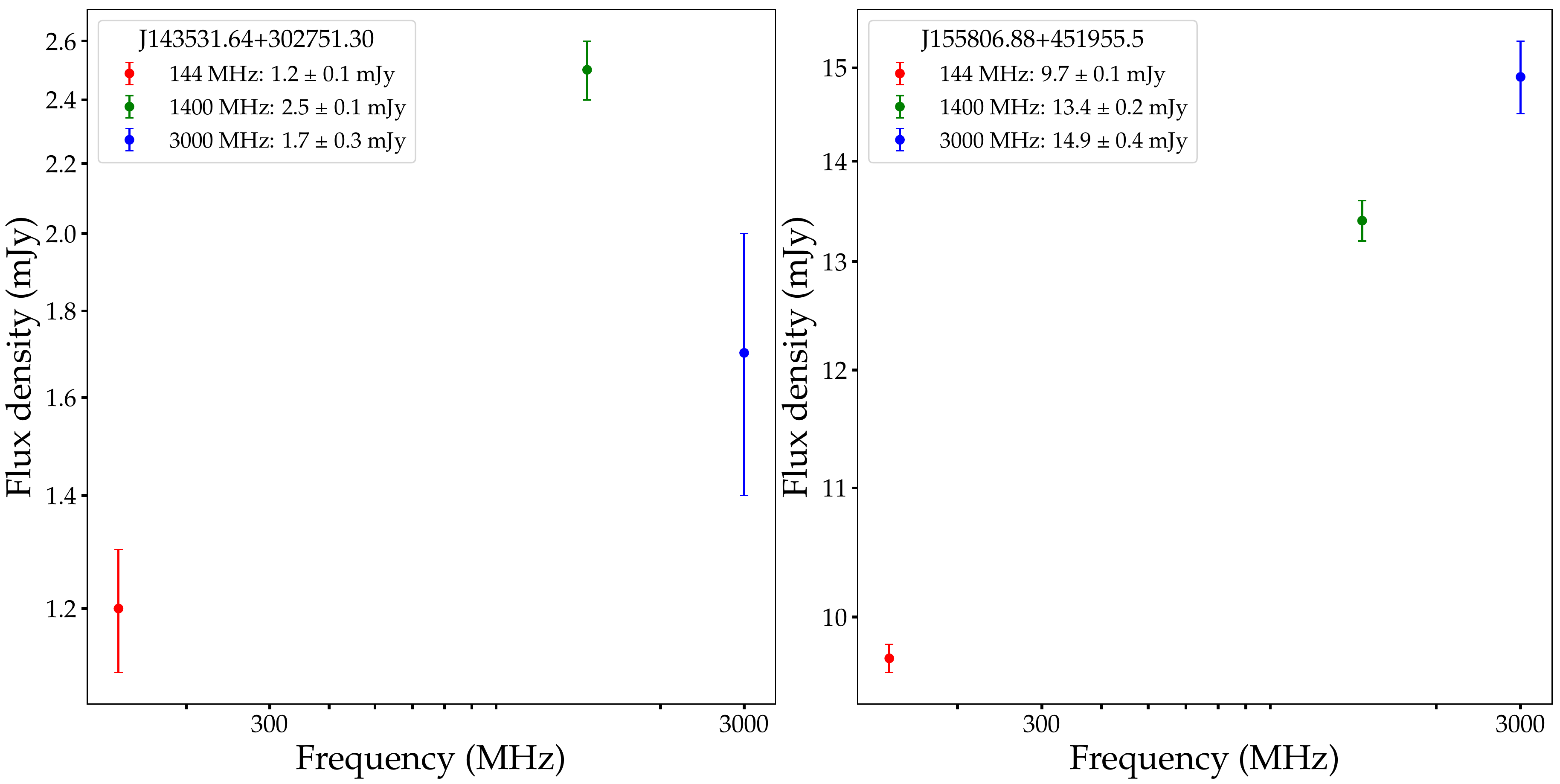}
  \caption{\label{fig:gpsplot} Radio spectra of two candidate Gigahertz Peaked Spectrum sources from our sample of  G-DDRGs as described in Sec.\ \ref{sec:gps}. Flux densities measured from LoTSS, FIRST, and VLASS are represented in red, green, and blue colours, respectively.}
\end{figure*}

\begin{figure*}
\centering
  \includegraphics[scale=0.35]{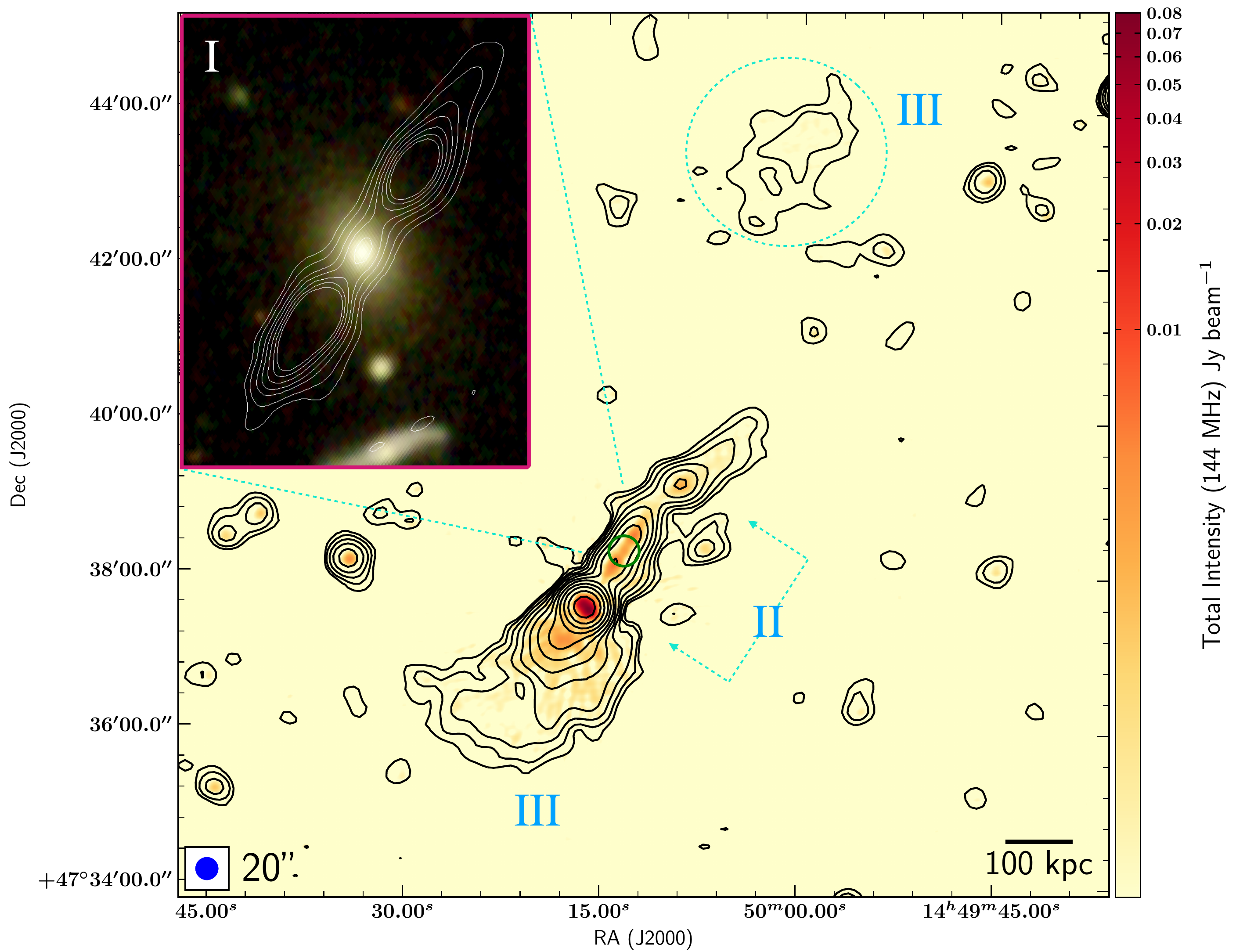}
  \caption{\label{fig:TDRGC} LoTSS DR2 144 MHz image with angular resolution of ~6$\arcsec$ in colour with ~20$\arcsec$ LoTSS contours overlaid (black contours). There are 15 linearly spaced contour levels between 0.000435 to 0.4995 Jy~beam$^{-1}$, where the first value is 3$\sigma$. In the top left inset figure an optical image is overlaid with a VLASS image with 2.5$\arcsec$ angular resolution. For more details see Sec.~\ref{sec:TDRG}. II and III markings indicate the second and outermost lobes.}
\end{figure*}

\begin{figure*}
  \includegraphics[scale=0.3]{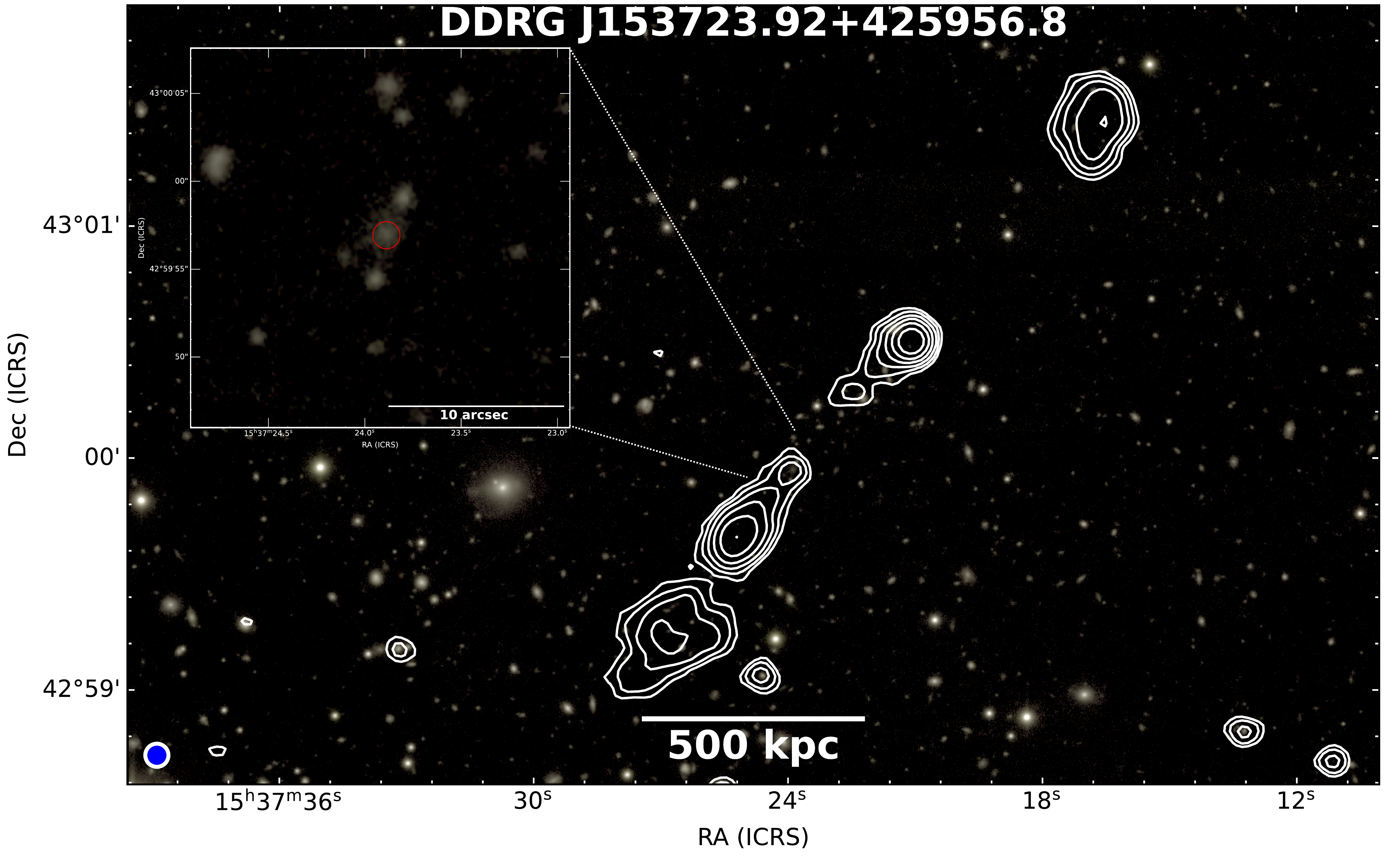}
  \caption{\label{fig:highz} HSC multi-colour image of the  G-DDRG J153723.92+425956.8, overlaid with LoTSS 6\arcsec~ contours in white (8 equispaced levels starting from 0.3 mJy~beam$^{-1}$). The inset in the top left corner shows the distant host galaxy ($z \sim$1.4) of the DDRG. For further details, see to Sec.~\ref{sec:env}.}
\end{figure*}

\subsection{Core spectral index} \label{sec:coreSI}
Using survey data from LoTSS at 144 MHz, FIRST at 1400 MHz \citep{FIRST2015Helfand}, and VLASS\footnote{\url{https://cirada.ca/vlasscatalogueql0}} \citep{Gordon2020VLASS} at 3000 MHz, we have measured the core flux densities for our sample sources. These measurements were performed based on the availability of data in each respective survey and the resolved structures detected in the sources. 
We were only able to resolve and identify radio cores for 31 sources in FIRST, 45 in VLASS, and 35 in LoTSS DR2 (high resolution) from the total sample. 
For those sources with a resolved core, we measured the flux densities and estimated the spectral index using two or three frequency points. These measurements are presented in Tab.~\ref{tab:2}.

Most of the sources show a flatter spectral index ($\alpha < 0.5$) for the radio core which is expected with a few exceptions.
The core spectral indices of G-DDRGs, ILTJ083201.90+395548.2, ILTJ092743.90+293232.4, and ILTJ145302.50+330918.1 are steep, with $\alpha \geq$\,0.9 (see Tab.~\ref{tab:2}), derived from measurements obtained from FIRST and VLASS. This steep spectral index likely results from the bridge emission between the core and the inner lobes, as indicated by the 5\arcsec~ resolution FIRST map, which does not have the spatial resolution necessary to distinguish between core and bridge emissions.
Several radio cores of sources exhibit signs of turnover in their radio spectra, characterised by negative spectral indices. This turnover can be attributed to mechanisms such as free-free absorption or synchrotron self-absorption. These phenomena are likely at play due to the compact nature of the radio cores and their interactions with the host galaxies. The dense ionized gas surrounding the cores can cause free-free absorption, while the high density of relativistic electrons and magnetic fields within the compact sources can lead to synchrotron self-absorption, both contributing to the observed spectral turnovers.

\subsection{Cores of DDRGs as GPS}\label{sec:gps}
Gigahertz Peaked Spectrum (GPS) sources are a class of RLAGN characterised by their compact size and a spectral peak in the gigahertz range, indicating they might be in an early stage of development or simply young radio sources. Their typical sizes are within 1 kpc, often contained within the central regions of their host galaxies. 
They are often found in galaxies with active star formation or interaction with other galaxies \citep[for a review, see][]{2021OdeaSaikiaRev}. It is thought that GPS sources could evolve into larger radio galaxies like DDRGs, especially if they undergo multiple episodes of jet activity. A case of GPS as the inner part of a DDRG has been reported by \cite{2007Saikia} for DDRG J1247+6723.
To examine this connection we considered the core flux densities of our DDRGs in our sample using LoTSS, FIRST and VLASS. We found 2 new GPS candidates based on our analysis: J143531.64+302751.30 and J155806.88+451955.5, with peak frequencies $\sim$\, 1.4 GHz and $>$\,3 GHz, respectively. Their respective radio-core spectra are presented in Fig.~\ref{fig:gpsplot}. Additionally, to confirm their nature, it will be necessary to conduct repeated multi-frequency observations over a period of time to monitor flux variability, and also measure their flux densities at higher frequencies. If confirmed, these sources may manifest potentially as `triple-double radio galaxies', which show three distinct sets of lobes indicating 3 epochs of AGN jet activity. Also, DDRG ILTJ114722.34+350106.0 or B1144+352 is already a known GPS \citep{Snellen1995,Schoenmakers1999GPS} and is part of our sample under study in this work.

\subsection{Triple double radio galaxies (TDRGs)} \label{sec:TDRG}
Only in the past two decades with advancement in sensitivities of radio maps with sufficient angular resolution, RGs displaying three distinct sets of radio lobes have been reported. This exceptional class of RGs is referred to as `triple-double' radio galaxies (TDRGs), and so far, only 4 such examples have been reported. These findings are summarised in \citet{Chavan2023}, which reported the fourth TDRG recently. Identifying TDRGs is challenging due to their unusual structures. The outermost or earliest lobes are often quite asymmetric and relic-like, as their activity has ceased and a significant amount of time has elapsed since their formation. This makes it difficult to establish their association with the innermost lobes. Only with deep radio maps, such as those provided by LoTSS, which offer sufficient sensitivity to detect low surface brightness features, can we reliably identify these outermost or earliest diffuse relic lobes.
In the current work, we have identified ILTJ145013.45+473818.6 as a candidate TDRG hosted by a BCG at z$\sim$\,0.11 with the total size of $\sim$\,1.1 Mpc. It is shown in Fig.~\ref{fig:TDRGC} with the three episodes marked in blue. The southeastern outermost (III) lobe, from the earliest episode, is not as distinct and appears embedded in the extension of the southeastern lobe from the II epoch. Both the II and III lobes display asymmetry, with the southern lobe being closer to the radio core. The two sets of activity in this restarted galaxy seem to have mitigated the asymmetric environment, enabling the growth of the innermost (I) lobe in a more pristine cocoon with a homogeneous environment.
Further, deeper radio observations are needed to map the outermost/earliest (III) episode lobes, as well as higher angular resolution observations to map the I and II lobes accurately. These observations are particularly crucial to ascertain the true nature of the bright emission, possibly a hotspot, towards the southeastern region. If confirmed, this will be the fifth reported case of a TDRG.

\subsection{Merger driven restarted activity?}\label{sec:merger}
Galaxy mergers can funnel large amounts of gas towards the SMBH, triggering intense nuclear accretion. This process is often accompanied by the formation of AGN-driven jets and outflows. The mechanical feedback from AGN activity, particularly through jets, can heat the surrounding IGM and expel gas from the galaxy \citep[e.g., see review by][and references therein]{2024Harrison}. This feedback mechanism is crucial in regulating the gas supply, which in turn affects subsequent star formation and AGN activity \citep[e.g.,][]{2022Murthy,2022Ruffa}. 

Observations of galaxies that have recently undergone mergers often show signs of AGN activity, such as enhanced accretion rates and the presence of powerful jets. These post-merger galaxies frequently exhibit morphological disturbances and tidal features, which are indicative of recent interactions and the resultant AGN-driven processes. Several studies have shown that RLAGNs are strongly associated with mergers \citep[e.g.,][]{Cristina2012,Chiaberge2015,Bernhard2022}.

A notable example of merger-driven restarted activity in a radio galaxy, specifically observed in DDRGs, is CGCG 292-057 \citep{2012koziel,2023MISRA}.
In order to ascertain whether similar evidence of mergers can be observed in DDRGs, we carried out an examination of the host galaxy images for all sources in our sample. It was conducted using the deeper optical maps available from the DESI Legacy Imaging Surveys, and where these were not accessible, data from the SDSS and Pan-STARRS were employed. Our analysis identified merger signatures, such as tidal tails or proximate companion galaxies, in only six of the sources, which are depicted in Fig.~\ref{fig:merger}. In the case of the companion galaxies, the redshifts (either Pz or Spz) are similar.
Such studies serve as an initial step towards evaluating the hypothesis proposed by \citet{2003Liu}, which suggests that DDRGs could be remnants of merged supermassive binary black holes. By investigating the connection between galaxy mergers and DDRG formation, these studies can provide insights into the dynamics of supermassive black hole mergers and their impact on the surrounding intergalactic medium. It is challenging to draw definitive conclusions from the currently available optical data. Deeper imaging of the entire sample is required to uniformly investigate evidence of past mergers, which could shed light on the coalescence of black holes and their potential link to the interruption of jet activity.

\section{Galaxy cluster environment of DDRGs} \label{sec:env}
Recent studies of \citet{SAGAN4} and \citet{2024Oei} have shown that about 25\% of GRGs reside in relatively denser environments of galaxy clusters.
To investigate the presence of DDRGs in dense environments such as galaxy clusters, we performed a cross-matching analysis of our sample with cluster catalogues \citep{Szabo2011,WH2015, WHY2018, WH2021}. This analysis determined whether the sources were associated with the brightest cluster galaxies (BCGs). We identified 15 sources from our sample as BCGs, as detailed in Tab.~\ref{tab:clusters}. This indicates that the host galaxies of these DDRGs are the most luminous within their respective clusters. The properties of these BCG-hosted DDRGs are presented in Tab.~\ref{tab:clusters}. This study provides the first substantial evidence of giant restarted radio galaxies located at the centres of galaxy clusters. Notably, two DDRGs are in massive galaxy clusters with masses (M$\rm _{500}$) greater than $\rm \sim2 \times 10^{14}~  M_{\odot}$, indicating their powerful jetted AGNs.

Another interesting finding is that of the G-DDRG
J153723.92+425956.8 (ILTJ153724.28+425952.5), which is at a high redshift of 1.4 (Fig.~\ref{fig:highz}) in a galaxy cluster. Although the redshift measurement is not spectroscopic, it can be deemed reliable as it is derived using 7-band photometric data of Hyper Suprime-Cam Subaru and the Wide-field Infrared Survey Explorer as given in \citet{WH2021}. The radio core is observed in both FIRST and VLASS, which coincides with a 25.1 magnitude faint galaxy (only seen in \textit{Suprime-Cam Subaru}, see inset in Fig.~\ref{fig:highz}). The same has been classified as AGN (WISEA J153723.88+425957.7) using WISE mid-infrared data  \citep{2018AssefWISE}.

\subsection{Cluster Weather-Induced Misalignment in DDRGs}
The simulation study by \citet{Mendygral2012} demonstrated that flow speeds of $\sim$\,400 \kms~ in galaxy clusters are sufficient to influence the morphologies of radio jets and lobes significantly. Their synthetic radio observations or simulations of intermittent jet activity, with a 50\% duty cycle, in a relatively relaxed galaxy cluster environment, resulted in DDRGs exhibiting wide-angle tail (WAT) morphologies. This specific simulation-based prediction has not been tested or observed so far. Remarkably, our findings show that three of our DDRGs closely match these predictions, providing empirical support for the simulation results and highlighting the significant impact of cluster dynamics on jet and lobe structures. As mentioned earlier in Sec.~\ref{sec:MA}, DDRGs J083201.85+395545.99 (ILTJ083201.90+395548.2), J131458.77+382708.7 (ILTJ131458.85+382709.1), and J163415.05+492842.8 (ILTJ163415.49+492846.4) exhibit significant misalignment between their inner and outer double lobes, resulting in structures resembling wide-angle tailed radio galaxies. Both these sources are hosted by BCGs (Tab.~\ref{tab:clusters}), where the dynamic cluster environment exacerbates the misalignment between the inner and outer lobes, leading to the observed non-co-aligned structures. This congruence underscores the utility of simulation studies in predicting and understanding the complex behaviours of radio galaxies in various environments. Our findings are consistent with other RG studies that have shown how cluster weather, including ram pressure and turbulence within the intracluster medium, can significantly alter jet paths and morphologies \citep[e.g.,][]{1998Burns,2012Gitti}. These interactions can lead to the bending, disruption, and reorientation of jets, particularly in BCGs where the gravitational potential and gas density are highest. The match between our observations and the simulation predictions provides a deeper understanding of how these dynamic processes shape the evolution of radio galaxies in clusters.

\section{Summary}
Using data from the LoTSS DR2 survey, supplemented with ancillary data from other radio and optical surveys, we have compiled a sample of 111 giant or megaparsec-scale DDRGs, with 76 of these being newly reported as DDRGs. This sample, the largest catalogue of DDRGs to date, provides statistically robust insights into the intermittent activity in RLAGNs. Notably, it includes 9 sources with z$>$1, extending DDRG studies to high redshifts. Studying DDRGs is crucial for understanding AGN duty cycles as they offer direct observational evidence of re-triggering and quiescence timescales. Our analysis, using a multitude of data sets, involved measuring key properties such as arm-length ratios, flux density ratios, and misalignment angles to better understand their structure and dynamics. These parameters were used to study the symmetry and stability of the jets and understanding the impact of the immediate environment on their growth and morphology. Specifically, we have shown that the cocoons in which the inner or the newer lobes grow are often contaminated as inferred from observed asymmetry in arm-length ratios of several sources. On the other hand observations of symmetric inner lobes show the presence of 
a more homogenous environment, possibly paved by the activity of the first epoch. Hence, such analyses are necessary for understanding the growth of RGs and their interaction with the environment.

A significant aspect of our study was the investigation of the large-scale environments of these DDRGs, particularly their association with galaxy clusters, a focus that has not been systematically studied before. By examining the dynamic conditions within these clusters, we have shown how such environments exacerbate the misalignment between the inner doubles and outer doubles, leading to the observed non-co-aligned structures. Our findings highlight that cluster weather, including ram pressure and turbulence, can play a substantial role in shaping morphologies of DDRGs. Three sources, which exhibit high misalignments and are hosted by BCGs, display WAT-like morphologies. This observation aligns with simulation studies that predict such scenarios. The empirical support for simulation results, particularly those predicting WAT morphologies in DDRGs, underscores the importance of combining observational data with theoretical models to advance our knowledge of radio galaxy formation and behaviour.

Furthermore, we explored whether the host galaxies of these DDRGs are undergoing mergers, utilising optical data to detect signs of such interactions. This examination provided additional context for the observed jet morphologies and their stability.
In addition to the primary findings, we identified two gigahertz peaked-spectrum candidates within the G-DDRG sample. These GPS sources are characterised by their compactness and high-frequency turnover, indicating young or recently restarted AGN activity. Moreover, we identified one triple-double radio galaxy candidate, a rare structure with only four other known instances, further contributing to our understanding of complex jet dynamics.

One of the aims of this study was also to determine the fraction of GRGs that exhibit clear signs of restarted activity, indicated by DDRG morphology. Quantifying this is challenging due to the absence of a complete sample of GRGs and the limited availability of radio surveys with very high angular resolution ($\lesssim$1\arcsec). However, using the GRG catalogue of \citet{oei2023}, we found that fewer than 3\% of giant radio galaxies show DDRG morphologies at arcsecond resolutions, which is the best available resolution to us at present. This statistic, previously unreported, underscores the rarity of DDRGs. Even with more robust and complete samples, the percentage is unlikely to change significantly, suggesting that DDRGs are indeed quite rare.

The existence of DDRGs confirms the idea of duty cycles in AGN activity, where periods of activity are interspersed with periods of quiescence. Studies from the deep LOFAR surveys of the Lockman Hole field \citep{Jurlin2020,2023Jurlin} conclude that restarted RGs are more commonly observed than their remnant counterparts. It highlights that the inactive phase preceding the restarting of these galaxies is relatively short, lasting only a few tens of millions of years \citep[see table 7 of][]{Marecki2021}. This brief dormant period, set against the extensive lifespan of RGs, underscores the dynamic and recurring nature of their life cycles. 
This result, combined with our findings, suggests that the relative rarity of DDRGs could be due to the shorter dormancy period in most restarted RGs or quasi-continuous activity in most RGs. Consequently, many restarted RGs do not exhibit the characteristic DDRG morphology. It appears that DDRGs are those restarted RGs that take longer than usual to restart their radio jet activity. This is also supported by spectral ageing estimates of a few DDRGs \citep[e.g.,][]{Konar2006,Jamrozy2007,Konar2013}, which reveal a significant age difference between the inner and outer lobes. In other words, for the double-double structures to be observed, the on-off period of the radio AGN must be sufficiently long to allow the lobes from the first epoch to travel far enough from the host before new lobes form, appearing as distinct inner lobes.

Automated methods (e.g., machine learning) have significantly expedited the process of identifying larger samples of radio galaxies \citep[e.g.,][]{Tang2022,2024Mostert}. These methods are continually improving, extending their capabilities to more complex cases, such as X-shaped or DDRGs. Their importance is already growing rapidly, driven by the increasing volume of data from ongoing deep and wide-area radio surveys. This trend is expected to accelerate with data from upcoming facilities like the Square Kilometre Array (SKA\footnote{\url{https://skao.int/}}), where manual analysis will no longer be feasible, making these advanced techniques indispensable for uncovering sources with complex morphologies.

By presenting the properties of G-DDRGs and their implications, our research provides valuable insights into the environmental factors and dynamic processes that govern the evolution of giant radio galaxies with signs of restarted activity. 
In summary, DDRGs are a unique class of restarted radio galaxies offering insights into the mechanisms of jet activity in AGN. Their study, particularly through large-scale surveys like LoTSS, contributes significantly to understanding galaxy evolution and the factors influencing AGN jet cycles.

%%%%%%%%%%%%%%%%%%%%%%%%%%%%%%%%%%%%%%%%%%%%%%%%%%%%%%%%%%%%%%%%%%%%%%%%%%%%%%%%%%%%%%%%%%%%%
\begin{acknowledgements}
We thank the anonymous reviewer for several comments and suggestions which have helped improve the manuscript.
PD acknowledges support from the Spanish Ministry of Science and Innovation under the grant -``PARSEC: Multiwavelength investigations of the central parsec of galaxies" (PID2020-114092GB-I00). LOFAR data products were provided by the LOFAR Surveys Key Science Project (LSKSP; \url{https://lofar-surveys.org/}) and were derived from observations with the International LOFAR Telescope (ILT). LOFAR \citep{vanHaarlem2013} is the Low Frequency Array designed and constructed by ASTRON. It has observing, data processing, and data storage facilities in several countries, which are owned by various parties (each with their own funding sources), and which are collectively operated by the ILT foundation under a joint scientific policy. The efforts of the LSKSP have benefited from funding from the European Research Council, NOVA, NWO, CNRS-INSU, the SURF Co-operative, the UK Science and Technology Funding Council and the J\"{u}lich Supercomputing Centre. We acknowledge the use of the DESI Legacy Imaging Surveys (\url{https://www.legacysurvey.org/acknowledgment/}). This research has made use of the CIRADA cutout service at \url{cutouts.cirada.ca}, operated by the Canadian Initiative for Radio Astronomy Data Analysis (CIRADA). CIRADA is funded by a grant from the Canada Foundation for Innovation 2017 Innovation Fund (Project 35999), as well as by the Provinces of Ontario, British Columbia, Alberta, Manitoba and Quebec, in collaboration with the National Research Council of Canada, the US National Radio Astronomy Observatory and Australia’s Commonwealth Scientific and Industrial Research Organisation. This research has made use of the NASA/IPAC Extragalactic Database (NED), which is funded by the National Aeronautics and Space Administration and operated by the California Institute of Technology. This research has made use of the VizieR catalogue tool, CDS, Strasbourg, France \citep{vizier}. We acknowledge that this work has made use of  \textsc{astropy} \citep{astropy}, \textsc{aplpy} \citep{apl}, and \textsc{topcat} \citep{top05}.

\end{acknowledgements}
%%%%%%%%%%%%%%%%%%%% REFERENCES %%%%%%%%%%%%%%%%%%
\bibliographystyle{aa} 
\bibliography{DDRG_LoTSSDR2}
%%%%%%%%%%%%%%%%%%%%%%%%%%%%%%%%%%%%%%%%%%%%%%%%%%

% %%%%%%%%%%%%%%%%% APPENDICES %%%%%%%%%%%%%%%%%%%%%
\appendix
\section{Tables} \label{sec:appendixA}

%%%%%%%%%%%%%%%%%%%%%%%%%%%%%%%%% Table 1 %%%%%%%%%%%%%%%%%%%%%%%%%%%%%

\onecolumn
\setlength{\tabcolsep}{2.35pt}
\begin{small}
\begin{longtable}{l c c c c c c c c c c c c c}

\captionsetup{width=\textwidth}

\caption{Properties of the 76 newly identified giant DDRGs or G-DDRGs from LoTSS DR2 in the upper half and in the lower half 35 G-DDRGs from literature in the LoTSS DR2 sky area. Column (2) lists the ILT name of the sources from the main LoTSS catalogue of \citet{Shimwell2022}.
Columns (3)- RA $\&$ (4)-Dec are the Right Ascension $\&$ Declination (J2000.0 
epoch) of the host galaxies of the DDRGs identified from one of the three surveys: SDSS, Pan-STARRS, or DESI Legacy Imaging Surveys. 
Column (5) is the redshift where the `$\dagger$' represents spectroscopic measurements. $\dagger^a$ in column (5) represents redshift from \citet{Cotter1996}.
Column (6) $\&$ (7) represent projected sizes for outer and inner lobes, respectively.
Columns (8),(9) and (10) represent the misalignment angle between outer lobes with core, the misalignment angle between inner lobes with core, and the misalignment angle between outer lobes with inner lobes, respectively. 
Columns (11) $\&$ (12) are the arm length ratios for outer $\&$ inner lobes, respectively. 
Column (13) represents the direction of components in which ratios are taken. We have taken the ratio of measured parameters for farther lobes to closer lobes from the host for the inner double. The same sense of direction is used for all the symmetry parameters as given in Column (13).
Flux densities and angular sizes are measured from the LoTSS DR2 radio maps. 
Superscripts in column (1) represent the parent sample from which the source has been taken.
$\mathsection$: \citet{Hardcastle2023}, *: \citet{oei2023}, and from literature parent sample- (a): \citet{Kozielrogue2020}, (b): \citet{PDLOTSS}, (c): \citet{Mahatma2019}, (d): \citet{Nandi2012}, (e): \citet{Renteria2017}, (f): \citet{Saikia2006}, (g): \citet{Schoenmakers1999GPS}, (h): \citet{SchoenmakersDDRG1},  (i): \citet{sagan1}, (j): \citet{Simonte2022}.
} \label{tab:main} \\

\hline
Sr.No &ILT names &               R.A &          Dec &                              Redshift &                            Outer size  & Inner size & MA$_{\rm O}$ &MA$_{\rm I}$ &MA$_{\rm O-I}$ &$\rm R_{\theta (o)}$ &$\rm R_{\theta (in)}$ & Dir. \\
- & - &  (HMS) & (DMS)  & ($z$) & (\arcsec/kpc) & (\arcsec/kpc) & (deg) & (deg) & (deg) & - & - & -\\
 (1) & (2) & (3) & (4) & (5) & (6) & (7) & (8) & (9) & (10) & (11) & (12) & (13) \\
\hline
\endfirsthead
\caption{continued.}\\
\hline\hline
SrNo &ILT names &               RA &          Dec &                              Redshift &                            Outer size  & Inner size & MA$_{\rm O}$ &MA$_{\rm I}$ &MA$_{\rm O-I}$ &$\rm R_{\theta (o)}$ &$\rm R_{\theta (in)}$ & Dir. \\
- & - &  (HMS) & (DMS)  & ($z$) & (\arcsec/kpc) & (\arcsec/kpc) & (deg) & (deg) & (deg) & - &- & -\\
 (1) & (2) & (3) & (4) & (5) & (6) & (7) & (8) & (9) & (10) & (11) & (12) & (13) \\
\hline
\endhead
\hline
\endfoot
1$^{*}$ &ILTJ000345.85+313924.7 &00 03 45.91 &31 39 24.24 &0.5548$\dagger$ &325 (2154) &144 (953) &0.7 &4.3 &5.6 &1.4 &1.3 &E-W \\
2$^{\mathsection}$ &ILTJ014149.55+274045.3 &01 41 49.60 &27 40 45.7 &0.2560 &268 (1066) &16 (63) &14.8 &0.8 &1.5 &1.0 &1.5 &N-S \\
3$^{\mathsection}$ &ILTJ024023.22+272929.4 &02 40 23.20 &27 29 28.1 &0.2030 &219 (731) &16 (55) &11.6 &2.8 &14.6 &2.0 &1.2 &N-S \\
4$^{*}$ &ILTJ081538.31+495559.6 &08 15 38.42 &49 56 00.30 &0.5962$\dagger$ &148 (1018) &22 (152) &21.9 &10.6 &7.0 &2.0 &1.3 &N-S \\
5$^{*}$ &ILTJ083110.34+443154.6 &08 31 10.57 &44 31 54.22 &0.3705$\dagger$ &206 (1085) &61 (319) &6.5 &11.4 &5.0 &1.4 &1.8 &W-E \\
6$^{*}$ &ILTJ083201.90+395548.2 &08 32 01.85 &39 55 45.99 &0.3195$\dagger$ &247 (1182) &18 (89) &67.1 &31.4 &3.0 &1.2 &1.1 &W-E \\
7$^{*}$ &ILTJ083614.75+344948.1 &08 36 14.28 &34 49 45.48 &0.4536$\dagger$ &341 (2035) &46 (271) &7.1 &22.5 &0.8 &0.8 &1.9 &S-N \\
8$^{*}$ &ILTJ084127.07+554626.6 &08 41 27.03 &55 46 27.18 &0.7912$\dagger$ &406 (3120) &200 (1535) &2.5 &4.0 &0.6 &1.4 &2.4 &N-S \\
9$^{*}$ &ILTJ084458.37+420438.5 &08 44 58.39 &42 04 38.61 &0.1493$\dagger$ &416 (1118) &17 (45) &0.3 &7.6 &28.5 &0.9 &1.4 &W-E \\
10$^{*}$ &ILTJ090139.18+460757.0 &09 01 39.23 &46 07 56.38 &0.3539$\dagger$ &396 (2030) &57 (294) &3.5 &7.3 &5.1 &1.0 &1.2 &N-S \\
11$^{*}$ &ILTJ090440.20+375727.6 &09 04 39.27 &37 57 40.74 &0.3074$\dagger$ &205 (955) &47 (220) &0.3 &0.1 &3.2 &1.0 &1.3 &N-S \\
12$^{*}$ &ILTJ090815.36+320922.1 &09 08 15.63 &32 09 21.19 &0.4561$\dagger$ &299 (1786) &42 (249) &10.3 &21.2 &4.4 &1.5 &2.1 &E-W \\
13$^{*}$ &ILTJ091014.60+292100.7 &09 10 14.73 &29 21 19.09 &0.7321 &236 (1763) &39 (294) &2.0 &4.8 &0.8 &1.3 &1.1 &N-S \\
14$^{*}$ &ILTJ091303.87+511014.6 &09 13 05.16 &51 10 26.23 &0.2750 &395 (1705) &99 (427) &17.4 &22.5 &1.9 &1.7 &1.7 &N-S \\
15$^{*}$ &ILTJ091332.15+465817.6 &09 13 27.15 &46 58 14.43 &0.272$\dagger$ &171 (732) &34 (146) &26.1 &5.2 &4.7 &0.6 &1.3 &E-W \\
16$^{\mathsection}$ &ILTJ091432.12+670454.0 &09 14 31.90 &67 04 52.8 &0.7930 &128 (955) &21 (159) &2.6 &5.2 &2.1 &1.1 &1.4 &S-N \\
17$^{*}$ &ILTJ091623.04+484441.6 &09 16 22.31 &48 44 32.06 &0.4700 &260 (1582) &29 (177) &1.5 &3.5 &8.4 &0.9 &1.4 &N-S \\
18$^{\mathsection}$ &ILTJ092743.90+293232.4 &09 27 44.00 &29 32 34.3 &0.5930 &114 (756) &24 (160) &7.8 &6.7 &0.3 &1.3 &1.6 &W-E \\
19$^{*}$ &ILTJ094112.07+554659.3 &09 41 11.98 &55 46 59.78 &0.3084$\dagger$ &293 (1369) &12 (55) &22.7 &1.0 &2.0 &1.0 &1.2 &E-W \\
20$^{*}$ &ILTJ094825.70+512623.7 &09 48 26.27 &51 26 53.17 &0.459$\dagger$ &191 (1148) &35 (207) &16.9 &10.8 &8.3 &1.1 &1.5 &N-S \\
21$^{*}$ &ILTJ095341.71+654500.3 &09 53 41.89 &65 45 00.72 &0.2680 &240 (1019) &49 (206) &10.7 &5.3 &2.6 &0.7 &1.3 &E-W \\
22$^{\mathsection}$ &ILTJ100014.11+504536.6 &10 00 15.48 &50 44 56.9 &0.399$\dagger$ &148 (820) &35 (194) &8.0 &2.7 &2.1 &1.0 &1.2 &N-S \\
23$^{\mathsection}$ &ILTJ100410.75+505849.5 &10 04 10.68 &50 58 49.8 &0.2690 &214 (882) &63 (259) &10.3 &4.2 &0.5 &1.3 &1.5 &S-N \\
24$^{\mathsection}$ &ILTJ100956.68+365022.4 &10 09 56.00 &36 50 22.6 &0.7320 &117 (851) &39 (283) &30.3 &3.7 &36.6 &1.5 &1.2 &W-E \\
25$^{*}$ &ILTJ101352.64+451042.3 &10 13 52.54 &45 10 42.56 &0.5200 &169 (1083) &48 (311) &8.7 &4.1 &1.4 &0.8 &1.1 &E-W \\
26$^{*}$ &ILTJ102536.42+382137.8 &10 25 35.94 &38 21 48.40 &0.5130 &347 (2213) &33 (209) &7.2 &1.6 &0.6 &1.1 &1.7 &N-S \\
27$^{*}$ &ILTJ105415.68+341739.9 &10 54 15.84 &34 17 24.57 &0.7300 &216 (1613) &48 (357) &1.9 &3.1 &1.8 &0.7 &1.5 &N-S \\
28$^{\mathsection}$ &ILTJ110339.59+455006.1 &11 03 39.42 &45 50 10.5 &0.3050 &178 (803) &20 (90) &9.6 &1.1 &6.4 &1.3 &1.4 &S-N \\
29$^{\mathsection}$ &ILTJ111105.07+310100.8 &11 11 04.65 &31 01 01.2 &0.6470 &271 (1877) &30 (206) &3.7 &1.0 &1.5 &1.0 &1.1 &E-W \\
30$^{*}$ &ILTJ111449.37+394845.0 &11 14 49.23 &39 48 48.25 &0.5920 &190 (1297) &34 (234) &1.7 &2.9 &3.1 &0.9 &2.0 &N-S \\
31$^{\mathsection}$ &ILTJ112421.10+300408.9 &11 24 21.36 &30 03 59.5 &0.8255$\dagger$ &194 (1473) &23 (175) &1.8 &3.6 &1.4 &0.9 &1.1 &S-N \\
32$^{*}$ &ILTJ112625.32+645222.8 &11 26 24.90 &64 52 36.86 &0.6820 &239 (1738) &69 (505) &2.4 &6.3 &4.1 &0.8 &1.3 &N-S \\
33$^{*}$ &ILTJ114142.93+364605.4 &11 41 42.87 &36 46 05.23 &0.2651$\dagger$ &514 (2161) &53 (222) &8.7 &3.1 &7.8 &1.0 &1.4 &W-E \\
34$^{*}$ &ILTJ115930.96+332823.0 &11 59 31.17 &33 28 26.37 &0.605$\dagger$ &285 (1968) &24 (163) &17.3 &4.3 &0.2 &0.9 &1.3 &S-N \\
35$^{*}$ &ILTJ120753.01+532358.2 &12 07 52.44 &53 23 55.72 &0.5056$\dagger$ &233 (1476) &45 (283) &0.6 &3.8 &2.0 &1.2 &1.3 &E-W \\
36$^{*}$ &ILTJ121001.94+440524.6 &12 10 01.94 &44 05 24.38 &1.5660 &378 (3289) &99 (861) &2.6 &3.5 &1.2 &1.3 &1.1 &N-S \\
37$^{\mathsection}$ &ILTJ121534.59+591129.5 &12 15 36.37 &59 10 51.7 &0.5368$\dagger$ &120 (759) &49 (308) &11.2 &5.0 &1.1 &0.8 &1.2 &N-S \\
38$^{\mathsection}$ &ILTJ121843.18+550612.7 &12 18 43.11 &55 06 12.9 &0.2170 &307 (1081) &97 (340) &1.4 &10.4 &3.0 &1.4 &1.3 &N-S \\
39$^{\mathsection}$ &ILTJ123829.41+454655.5 &12 38 29.44 &45 46 53.3 &0.581$\dagger$ &138 (906) &20 (129) &7.0 &3.6 &1.1 &1.3 &1.1 &N-S \\
40$^{\mathsection}$ &ILTJ131458.85+382709.1 &13 14 58.77 &38 27 08.7 &0.643$\dagger$ &102 (702) &8 (53) &58.1 &20.8 &10.7 &0.9 &1.1 &E-W \\
41$^{\mathsection}$ &ILTJ131605.93+271951.8 &13 16 06.72 &27 19 27.9 &0.2497$\dagger$ &394 (1538) &16 (62) &14.3 &4.5 &6.5 &1.0 &2.6 &N-S \\
42$^{*}$ &ILTJ132110.00+362622.5 &13 21 10.07 &36 26 27.43 &0.2666$\dagger$ &234 (990) &12 (49) &23.0 &8.4 &7.8 &1.7 &1.1 &S-N \\
43$^{*}$ &ILTJ132227.31+613251.7 &13 22 27.37 &61 32 51.74 &0.5650 &289 (1936) &19 (126) &3.1 &8.1 &0.5 &2.2 &3.3 &N-S \\
44$^{*}$ &ILTJ134436.36+291239.4 &13 44 36.34 &29 12 39.61 &0.7660 &387 (2940) &108 (825) &2.9 &0.5 &2.9 &1.1 &1.1 &E-W \\
45$^{*}$ &ILTJ134713.30+240613.6 &13 47 12.54 &24 07 09.49 &0.6740 &175 (1265) &23 (166) &1.8 &3.1 &0.1 &0.4 &1.2 &S-N \\
46$^{\mathsection}$ &ILTJ134722.14+582019.2 &13 47 21.11 &58 20 19.4 &1.0110 &232 (1861) &27 (217) &3.1 &7.1 &0.1 &1.0 &1.2 &E-W \\
47$^{\mathsection}$ &ILTJ135847.29+323549.1 &13 58 47.11 &32 35 41.8 &0.5170 &284 (1814) &90 (575) &4.3 &17.9 &0.3 &1.1 &1.7 &S-N \\
48$^{*}$ &ILTJ141926.78+662731.9 &14 19 26.60 &66 27 30.02 &0.3360 &249 (1236) &17 (82) &7.9 &11.2 &6.1 &0.8 &2.0 &N-S \\
49$^{*}$ &ILTJ142015.30+373414.4 &14 20 14.04 &37 34 06.52 &0.5813$\dagger$ &233 (1582) &40 (272) &0.2 &0.0 &3.1 &1.0 &1.3 &S-N \\
50$^{\mathsection}$ &ILTJ142334.04+655838.0 &14 23 34.23 &65 58 40.2 &0.8360 &112 (854) &13 (102) &24.8 &31.1 &14.4 &1.5 &1.1 &S-N \\
51$^{*}$ &ILTJ142712.95+441720.3 &14 27 12.91 &44 17 20.36 &0.6550 &314 (2244) &55 (392) &13.4 &5.0 &5.2 &0.4 &1.2 &N-S \\
52$^{*}$ &ILTJ143431.12+652025.1 &14 34 32.77 &65 20 28.26 &0.7470 &222 (1673) &59 (442) &1.6 &0.1 &8.1 &0.9 &1.1 &E-W \\
53$^{*}$ &ILTJ143532.42+302757.0 &14 35 31.64 &30 27 51.30 &0.6716$\dagger$ &223 (1611) &72 (522) &2.4 &0.6 &1.4 &0.8 &1.2 &W-E \\
54$^{*}$ &ILTJ144609.52+445422.5 &14 46 09.22 &44 54 18.69 &0.2550 &268 (1094) &31 (125) &0.4 &2.6 &0.5 &1.2 &1.2 &N-S \\
55$^{*}$ &ILTJ150302.98+572403.3 &15 03 01.43 &57 23 18.70 &0.3770 &218 (1165) &43 (231) &1.3 &10.2 &2.3 &0.9 &1.9 &S-N \\
56$^{*}$ &ILTJ150547.75+392528.9 &15 05 47.79 &39 25 29.20 &0.6550 &269 (1926) &56 (402) &11.2 &6.1 &2.7 &1.5 &2.6 &S-N \\
57$^{\mathsection}$ &ILTJ152527.43+413215.5 &15 25 27.44 &41 32 13.3 &0.7843$\dagger$ &149 (1107) &25 (188) &27.9 &7.3 &3.5 &1.5 &1.2 &N-S \\
58$^{\mathsection}$ &ILTJ152756.90+531715.6 &15 27 57.11 &53 17 12.0 &1.1660 &110 (907) &13 (105) &10.7 &9.5 &4.6 &0.8 &1.1 &N-S \\
59$^{\mathsection}$ &ILTJ152949.12+491557.2 &15 29 49.69 &49 15 37.6 &1.0440 &125 (1009) &29 (238) &4.8 &8.8 &1.4 &0.9 &1.0 &N-S \\
60$^{*}$ &ILTJ153659.30+310539.5 &15 36 59.17 &31 05 39.0 &0.1992$\dagger$ &258 (876) &39 (132) &4.7 &4.9 &1.9 &0.9 &2.1 &W-E \\
61$^{\mathsection}$ &ILTJ153724.28+425952.5 &15 37 23.92 &42 59 56.8 &1.4015 &178 (1538) &69 (596) &4.8 &3.6 &1.7 &2.3 &2.0 &N-S \\
62$^{*}$ &ILTJ154952.31+555120.7 &15 49 52.28 &55 51 21.03 &0.6510 &193 (1375) &74 (526) &4.3 &1.8 &8.9 &1.2 &1.0 &E-W \\
63$^{*}$ &ILTJ155054.46+473449.3 &15 50 54.50 &47 34 49.54 &0.551$\dagger$ &180 (1187) &27 (178) &6.2 &4.8 &12.6 &1.4 &1.2 &E-W \\
64$^{\mathsection}$ &ILTJ155436.64+533430.1 &15 54 36.64 &53 34 30.1 &0.7608$\dagger$ &130 (956) &15 (111) &13.6 &2.7 &22.0 &1.0 &1.4 &S-N \\
65$^{\mathsection}$ &ILTJ155805.81+451958.7 &15 58 06.88 &45 19 55.5 &0.4180 &144 (793) &79 (435) &16.1 &3.4 &6.6 &1.0 &1.5 &N-S \\
66$^{\mathsection}$ &ILTJ161010.04+455552.7 &16 10 10.11 &45 55 56.1 &0.7094$\dagger$ &114 (817) &9 (62) &4.6 &2.3 &2.7 &1.4 &1.0 &S-N \\
67$^{\mathsection}$ &ILTJ163415.49+492846.4 &16 34 15.05 &49 28 42.8 &0.517$\dagger$ &139 (866) &35 (215) &64.3 &37.1 &26.1 &1.8 &1.7 &E-W \\
68$^{*}$ &ILTJ175253.19+372351.1 &17 52 53.46 &37 23 57.44 &0.3150 &212 (1006) &104 (494) &8.1 &1.9 &0.7 &0.8 &1.2 &S-N \\
69$^{*}$ &ILTJ222719.14+221437.8 &22 27 19.10 &22 14 39.20 &0.174$\dagger$ &332 (1010) &12 (35) &1.3 &1.7 &7.6 &0.6 &1.2 &S-N \\
70$^{*}$ &ILTJ230341.84+253002.7 &23 03 41.81 &25 30 02.39 &0.9020 &289 (2315) &96 (766) &1.6 &1.9 &0.6 &1.5 &1.6 &S-N \\
71$^{*}$ &ILTJ231133.28+325119.3 &23 11 39.36 &32 52 02.32 &0.43$\dagger$ &294 (1698) &39 (227) &1.4 &2.2 &2.0 &0.9 &1.7 &E-W \\
72$^{*}$ &ILTJ231421.57+192339.3 &23 14 19.63 &19 23 25.93 &0.0924$\dagger$ &518 (919) &121 (215) &14.8 &16.8 &1.4 &0.8 &1.1 &E-W \\
73$^{\mathsection}$ &ILTJ232036.32+290056.0 &23 20 36.72 &29 00 56.5 &1.0090 &141 (1134) &24 (194) &17.0 &6.3 &4.3 &1.0 &1.5 &W-E \\
74$^{\mathsection}$ &ILTJ232642.14+184221.9 &23 26 41.50 &18 42 06.8 &1.0620 &112 (906) &9 (76) &17.7 &16.0 &13.0 &1.9 &1.6 &S-N \\
75$^{\mathsection}$ &ILTJ234901.19+231733.0 &23 49 01.22 &23 17 34.1 &0.9310 &204 (1603) &22 (170) &1.8 &5.5 &17.2 &1.0 &1.0 &N-S \\
76$^{\mathsection}$ &ILTJ235751.02+332610.1 &23 57 51.66 &33 26 08.2 &0.7390 &108 (789) &24 (175) &13.7 &20.2 &1.4 &1.0 &1.4 &S-N \\
\hline \\
1$^{f}$ &ILTJ004145.91+322453.1 &00 41 46.12 &32 24 52.65 &0.4300 &165 (956) &31 (180) &1.4 &3.1 &3.0 &1.5 &1.2 &W-E \\
2$^{a}$ &ILTJ084525.79+522916.4 &08 45 25.52 &52 29 15.7 &0.4029$\dagger$ &231 (1247) &25 (134) &4.4 &0.4 &0.1 &0.9 &1.2 &E-W \\
3$^{c}$ &ILTJ105133.87+514449.9 &10 51 34.43 &51 44 55.28 &0.8710 &121 (959) &24 (188) &5.6 &3.0 &0.7 &1.4 &1.1 &S-N \\
4$^{b}$ &ILTJ105742.51+510558.5 &10 57 43.09 &51 05 57.68 &0.4627$\dagger$ &135 (813) &54 (325) &0.5 &1.5 &3.5 &1.0 &1.4 &E-W \\
5$^{b}$ &ILTJ110457.23+480903.4 &11 04 57.07 &48 09 13.70 &0.4147$\dagger$ &142 (805) &49 (275) &10.2 &2.3 &0.2 &2.0 &1.4 &N-S \\
6$^{b}$ &ILTJ110613.66+485748.3 &11 06 13.55 &48 57 48.29 &0.8140 &246 (1909) &57 (444) &15.8 &3.4 &8.4 &0.5 &1.2 &N-S \\
7$^{c}$ &ILTJ112218.57+555037.0 &11 22 18.49 &55 50 33.83 &0.91$\dagger$ &96 (772) &24 (195) &9.5 &14.8 &9.3 &1.3 &1.7 &S-N \\
8$^{b}$ &ILTJ112425.26+554614.1 &11 24 25.08 &55 46 15.6 &0.809$\dagger$ &139 (1044) &47 (355) &31.2 &13.3 &2.6 &1.0 &1.1 &N-S \\
9$^{g}$ &ILTJ114722.34+350106.0 &11 47 22.13 &35 01 07.5 &0.0629$\dagger$ &726 (907) &43 (53) &9.9 &2.9 &11.7 &1.2 &1.6 &E-W \\
10$^{c}$ &ILTJ115528.85+485101.3 &11 55 28.25 &48 50 44.2 &1.0310 &148 (1193) &25 (199) &10.7 &23.1 &0.6 &1.1 &1.1 &S-N \\
11$^{c}$ &ILTJ120500.71+475843.6 &12 04 59.91 &47 58 27.63 &0.5650 &171 (1141) &27 (181) &4.8 &7.7 &5.4 &1.1 &1.3 &S-N \\
12$^{c}$ &ILTJ121502.39+474640.8 &12 15 2.25 &47 46 41.71 &0.597$\dagger$ &105 (720) &20 (134) &10.4 &3.2 &1.4 &1.3 &1.0 &E-W \\
13$^{b}$ &ILTJ121859.61+505249.8 &12 19 00.76 &50 52 54.41 &0.3851$\dagger$ &154 (835) &14 (74) &28.9 &2.6 &9.8 &0.6 &1.3 &N-S \\
14$^{c}$ &ILTJ122542.02+520006.1 &12 25 44.63 &51 59 51.75 &1.0140 &131 (1081) &33 (271) &1.5 &1.6 &8.3 &1.1 &1.1 &E-W \\
15$^{c}$ &ILTJ123005.45+491515.9 &12 30 05.47 &49 15 16.1 &0.9290 &185 (1491) &12 (99) &1.1 &10.3 &4.1 &0.8 &1.1 &W-E \\
16$^{h}$ &ILTJ124237.53+383802.6 &12 42 36.9 &38 38 06.3 &0.4077$\dagger$ &135 (756) &56 (314) &10.4 &1.1 &8.4 &0.9 &1.9 &W-E \\
17$^{c}$ &ILTJ124548.71+563109.3 &12 45 48.73 &56 31 11.77 &0.7230 &101 (750) &9 (70) &2.3 &5.0 &6.3 &1.2 &1.0 &S-N \\
18$^{e}$ &ILTJ130126.66+510458.1 &13 01 25.90 &51 05 0.70 &0.2690 &558 (2372) &34 (144) &2.7 &8.1 &0.3 &1.3 &3.1 &E-W \\
19$^{c}$ &ILTJ130354.23+464233.5 &13 03 57.86 &46 42 50.55 &0.5843$\dagger$ &179 (1218) &52 (356) &2.2 &5.1 &1.6 &1.1 &1.0 &W-E \\
20$^{b}$ &ILTJ134311.00+555940.1 &13 43 13.31 &56 00 08.35 &0.4848$\dagger$ &225 (1392) &93 (574) &8.9 &5.2 &1.6 &1.7 &1.7 &N-S \\
21$^{d}$ &ILTJ140719.34+513157.1 &14 07 18.48 &51 32 04.63 &0.3405$\dagger$ &200 (1000) &18 (92) &2.1 &4.7 &4.3 &0.9 &1.5 &W-E \\
22$^{b}$ &ILTJ140719.30+550605.3 &14 07 19.96 &55 06 01.29 &0.3322$\dagger$ &154 (757) &16 (78) &1.2 &4.8 &4.1 &1.1 &1.4 &E-W \\
23$^{b}$ &ILTJ141504.73+463429.1 &14 15 04.70 &46 34 28.97 &0.2240 &396 (1469) &96 (358) &38.4 &6.2 &3.8 &0.7 &1.4 &E-W \\
24$^{j}$ &ILTJ142335.95+330238.9 &14 23 38.54 &33 02 54.1 &0.9410 &105 (832) &25 (196) &1.6 &0.5 &1.9 &0.9 &1.2 &N-S \\
25$^{j}$ &ILTJ143652.33+341636.7 &14 36 53.17 &34 16 50.56 &0.2500 &217 (874) &61 (246) &10.5 &14.1 &0.9 &1.0 &2.4 &N-S \\
26$^{c}$ &ILTJ143735.71+514434.8 &14 37 37.66 &51 44 46.45 &0.8510 &163 (1282) &21 (166) &6.6 &6.1 &2.6 &1.3 &1.2 &E-W \\
27$^{h}$ &ILTJ145302.50+330918.1 &14 53 02.86 &33 08 42.40 &0.2481$\dagger$ &336 (1348) &37 (147) &10.2 &1.4 &7.4 &1.3 &1.1 &S-N \\
28$^{c}$ &ILTJ145609.66+482011.2 &14 56 11.25 &48 19 28.0 &0.8080 &138 (1071) &22 (171) &14.8 &8.3 &0.0 &1.0 &1.3 &S-N \\
29$^{c}$ &ILTJ145641.16+484940.5 &14 56 40.66 &48 49 42.73 &0.8270 &136 (1060) &21 (164) &6.9 &6.8 &1.4 &1.1 &1.5 &W-E \\
30$^{a}$ &ILTJ160811.30+325427.2 &16 08 10.8 &32 54 18.9 &0.3038$\dagger$ &160 (740) &31 (142) &26.4 &15.4 &5.5 &1.4 &1.5 &E-W \\
31$^{i}$ &ILTJ161242.66+431314.8 &16 12 42.06 &43 13 19.82 &0.2507$\dagger$ &251 (1013) &20 (82) &0.3 &12.3 &0.8 &1.6 &1.4 &N-S \\
32$^{i}$ &ILTJ161601.20+482535.9 &16 16 01.26 &48 25 35.40 &0.2330 &756 (2890) &76 (292) &0.3 &4.9 &6.8 &1.2 &1.6 &N-S \\
33$^{d}$ &ILTJ162754.57+290624.3 &16 27 54.64 &29 06 20.36 &0.7300 &96 (718) &10 (78) &5.8 &1.4 &4.4 &1.0 &2.0 &N-S \\
34$^{d}$ &ILTJ170517.88+394025.2 &17 05 17.84 &39 40 29.2 &0.8110 &221 (1665) &19 (141) &0.5 &10.4 &0.9 &1.6 &1.7 &S-N \\
35$^{d}$ &ILTJ170625.53+434037.4 &17 06 25.4 &43 40 40.19 &0.8500 &147 (1157) &32 (249) &12.2 &1.3 &0.6 &1.0 &1.3 &W-E \\

\hline
\end{longtable}
\end{small}
%%%%%%%%%%%%%%%%%%%%%%%

%%%%%%%%%%%%%%%% Table 2 %%%%%%%%%%%%%%%%%%%%%%%%%%

\setlength{\tabcolsep}{2.0pt}
\begin{small}
\begin{longtable}{l c c c c c c c c c c c c c}

\captionsetup{width=\textwidth}

\caption{Flux densities, flux density ratios and spectral indices for our sample of G-DDRGs. Superscripts in Column (1) have the same meaning as Tab.~\ref{tab:main}. Column (2) lists the ILT name of the sources from the main LoTSS catalogue of \citet{Shimwell2022}.
Columns (3) $\&$ (4) are the flux density ratios for outer $\&$ inner lobes, respectively. 
Column (5) represents the direction of components in which ratios are taken, same as Column (13) in \ref{tab:main}. Column (6) $\&$ (7) represents integrated flux density for the outer and inner lobes, respectively. Columns (8), (9), and (10) represent core flux density from LoTSS, FIRST, and VLASS, respectively. Column (11) represents either a two-point or three-point spectral index of the core based on available flux densities from Columns (8), (9), and (10). In Columns (8), (9), and (10), `$\times$' represents the radio source lying outside the sky coverage area, while `-' represents no detection or core not resolved from the extended emission in the respective radio sky survey. For the candidate GPS source ILTJ143532.42+302757.0 or J143531.64+302751.30, two spectral indices are given. } \label{tab:2} \\

\hline
SrNo & ILT~names &$\rm R_{S(o)}$ &$\rm R_{S(in)}$ & Dir. & $S_{144(Outer)}$ & $S_{144(Inner)}$ & $S_{144(core)}$ & $S_{1400(core)}$ & $S_{3000(core)}$ & $\alpha_{core}$ \\
& & & & & (mJy) & (mJy) & (mJy) & (mJy) & (mJy) & \\
 (1) & (2) & (3) & (4) & (5) & (6) & (7) & (8) & (9) & (10) & (11) \\
\hline
\endfirsthead
\caption{continued.}\\
\hline\hline
SrNo & ILT~names &$\rm R_{S(o)}$ &$\rm R_{S(in)}$ & Dir. & $S_{144(Outer)}$ & $S_{144(Inner)}$ & $S_{144(core)}$ & $S_{1400(core)}$ & $S_{3000(core)}$ & $\alpha_{core}$ \\
& & & & & (mJy) & (mJy) & (mJy) & (mJy) & (mJy) & \\
 (1) & (2) & (3) & (4) & (5) & (6) & (7) & (8) & (9) & (10) & (11) \\
\hline
\endhead
\hline
\endfoot
1$^{*}$ &ILTJ000345.85+313924.7 &0.7 &1.1 &E-W &226.8 $\pm$ 17.1 &133.8 $\pm$ 10.1 &5.4 $\pm$ 0.4 &X &4.5 $\pm$ 0.3 &0.06 $\pm$ 0.03 \\
2$^{\mathsection}$ &ILTJ014149.55+274045.3 &1.1 &1.2 &N-S &61 $\pm$ 4.6 &156 $\pm$ 11.7 &- &X &- &- \\
3$^{\mathsection}$ &ILTJ024023.22+272929.4 &0.9 &0.9 &N-S &35.4 $\pm$ 2.7 &20.3 $\pm$ 1.6 &- &X &- &- \\
4$^{*}$ &ILTJ081538.31+495559.6 &0.8 &0.7 &N-S &23.7 $\pm$ 1.8 &4.9 $\pm$ 0.4 &2.2 $\pm$ 0.3 &1.3 $\pm$ 0.1 &2.1 $\pm$ 0.5 &0.08 $\pm$ 0.18 \\
5$^{*}$ &ILTJ083110.34+443154.6 &1.7 &2.6 &W-E &71.9 $\pm$ 5.5 &13.3 $\pm$ 1.1 &- &- &- &- \\
6$^{*}$ &ILTJ083201.90+395548.2 &0.3 &1.2 &W-E &33.7 $\pm$ 2.6 &46.8 $\pm$ 3.5 &- &4.1 $\pm$ 0.1 &1.6 $\pm$ 0.1 &1.09 $\pm$ 0.14 \\
7$^{*}$ &ILTJ083614.75+344948.1 &0.9 &0.3 &S-N &32 $\pm$ 2.4 &89.6 $\pm$ 6.7 &- &- &1.1 $\pm$ 0.2 &- \\
8$^{*}$ &ILTJ084127.07+554626.6 &0.2 &1.3 &N-S &136.6 $\pm$ 10.3 &59.5 $\pm$ 4.5 &1.9 $\pm$ 0.3 &- &X &- \\
9$^{*}$ &ILTJ084458.37+420438.5 &1.7 &0.8 &W-E &36 $\pm$ 2.8 &73 $\pm$ 5.5 &- &- &- &- \\
10$^{*}$ &ILTJ090139.18+460757.0 &1.9 &0.6 &N-S &26 $\pm$ 2 &6.7 $\pm$ 0.6 &1.7 $\pm$ 0.3 &- &- &- \\
11$^{*}$ &ILTJ090440.20+375727.6 &0.6 &0.4 &N-S &17.8 $\pm$ 1.4 &40.5 $\pm$ 3.1 &- &- &- &- \\
12$^{*}$ &ILTJ090815.36+320922.1 &0.6 &0.1 &E-W &142.5 $\pm$ 10.8 &218.9 $\pm$ 16.5 &- &- &4.9 $\pm$ 0.2 &- \\
13$^{*}$ &ILTJ091014.60+292100.7 &1.4 &1.2 &N-S &226.3 $\pm$ 17.1 &65.7 $\pm$ 5 &- &- &- &- \\
14$^{*}$ &ILTJ091303.87+511014.6 &0.5 &0.4 &N-S &59.7 $\pm$ 4.5 &33.2 $\pm$ 2.5 &- &- &2 $\pm$ 0.2 &- \\
15$^{*}$ &ILTJ091332.15+465817.6 &1.7 &1.0 &E-W &21 $\pm$ 1.6 &3.2 $\pm$ 0.3 &1.4 $\pm$ 0.4 &- &- &- \\
16$^{\mathsection}$ &ILTJ091432.12+670454.0 &1.0 &0.5 &S-N &21.5 $\pm$ 1.7 &14.1 $\pm$ 1.1 &- &X &- &- \\
17$^{*}$ &ILTJ091623.04+484441.6 &1.3 &0.9 &N-S &100.1 $\pm$ 7.5 &7 $\pm$ 0.5 &0.5 $\pm$ 0.2 &- &- &- \\
18$^{\mathsection}$ &ILTJ092743.90+293232.4 &1.0 &0.8 &W-E &201.3 $\pm$ 15.1 &36.1 $\pm$ 2.7 &- &15.3 $\pm$ 0.2 &7.7 $\pm$ 0.2 &0.9 $\pm$ 0.03 \\
19$^{*}$ &ILTJ094112.07+554659.3 &0.6 &0.8 &E-W &33 $\pm$ 2.5 &14.4 $\pm$ 1.1 &- &- &- &- \\
20$^{*}$ &ILTJ094825.70+512623.7 &- &- &N-S &- &- &- &6.3 $\pm$ 0.1 &- &- \\
21$^{*}$ &ILTJ095341.71+654500.3 &2.7 &0.4 &E-W &8.8 $\pm$ 0.7 &2.7 $\pm$ 0.2 &2.7 $\pm$ 0.1 &X &- &- \\
22$^{\mathsection}$ &ILTJ100014.11+504536.6 &- &- &N-S &- &- &- &- &- &- \\
23$^{\mathsection}$ &ILTJ100410.75+505849.5 &- &- &S-N &- &- &- &- &- &- \\
24$^{\mathsection}$ &ILTJ100956.68+365022.4 &1.2 &0.2 &W-E &30.6 $\pm$ 2.3 &17.8 $\pm$ 1.4 &- &- &- &- \\
25$^{*}$ &ILTJ101352.64+451042.3 &0.9 &1.4 &E-W &14.1 $\pm$ 1.1 &9.6 $\pm$ 0.8 &2.8 $\pm$ 0.1 &0.6 $\pm$ 0.2 &0.8 $\pm$ 0.2 &0.6 $\pm$ 0.08 \\
26$^{*}$ &ILTJ102536.42+382137.8 &3.3 &0.5 &N-S &120.6 $\pm$ 9.1 &6.8 $\pm$ 0.5 &- &- &- &- \\
27$^{*}$ &ILTJ105415.68+341739.9 &0.6 &3.8 &N-S &36.9 $\pm$ 2.8 &42 $\pm$ 3.2 &- &- &- &- \\
28$^{\mathsection}$ &ILTJ110339.59+455006.1 &0.9 &2.1 &S-N &78.9 $\pm$ 6 &111 $\pm$ 8.3 &- &- &- &- \\
29$^{\mathsection}$ &ILTJ111105.07+310100.8 &0.9 &1.5 &E-W &85.7 $\pm$ 6.5 &72.4 $\pm$ 5.5 &- &- &- &- \\
30$^{*}$ &ILTJ111449.37+394845.0 &1.0 &4.8 &N-S &50.4 $\pm$ 3.8 &10.2 $\pm$ 0.8 &- &2.9 $\pm$ 0.1 &2.4 $\pm$ 0.3 &0.24 $\pm$ 0.16 \\
31$^{\mathsection}$ &ILTJ112421.10+300408.9 &1.5 &0.4 &S-N &22.4 $\pm$ 1.7 &2.1 $\pm$ 0.2 &- &- &- &- \\
32$^{*}$ &ILTJ112625.32+645222.8 &0.7 &3.9 &N-S &113.4 $\pm$ 8.5 &15.4 $\pm$ 1.2 &1.4 $\pm$ 0.2 &X &1.6 $\pm$ 0.3 &-0.05 $\pm$ 0.07 \\
33$^{*}$ &ILTJ114142.93+364605.4 &0.5 &1.1 &W-E &282.5 $\pm$ 21.2 &4.4 $\pm$ 0.4 &1.3 $\pm$ 0.1 &1.1 $\pm$ 0.2 &1.9 $\pm$ 0.2 &-0.08 $\pm$ 0.16 \\
34$^{*}$ &ILTJ115930.96+332823.0 &1.5 &1.3 &S-N &11.2 $\pm$ 0.9 &34.8 $\pm$ 2.6 &- &1.9 $\pm$ 0.2 &1.4 $\pm$ 0.2 &0.4 $\pm$ 0.17 \\
35$^{*}$ &ILTJ120753.01+532358.2 &0.6 &0.8 &E-W &31.8 $\pm$ 2.5 &30.1 $\pm$ 2.3 &- &- &- &- \\
36$^{*}$ &ILTJ121001.94+440524.6 &0.1 &1.9 &N-S &431.2 $\pm$ 32.3 &7.4 $\pm$ 0.6 &1.2 $\pm$ 0.2 &- &1 $\pm$ 0.1 &0.07 $\pm$ 0.07 \\
37$^{\mathsection}$ &ILTJ121534.59+591129.5 &- &- &N-S &- &- &- &- &- &- \\
38$^{\mathsection}$ &ILTJ121843.18+550612.7 &- &- &N-S &- &- &1.4 $\pm$ 0.2 &0.8 $\pm$ 0.1 &1 $\pm$ 0.2 &0.13 $\pm$ 0.12 \\
39$^{\mathsection}$ &ILTJ123829.41+454655.5 &- &- &N-S &- &- &- &- &- &- \\
40$^{\mathsection}$ &ILTJ131458.85+382709.1 &- &- &E-W &- &- &- &- &- &- \\
41$^{\mathsection}$ &ILTJ131605.93+271951.8 &0.9 &1.4 &N-S &103.9 $\pm$ 7.8 &60.4 $\pm$ 4.5 &- &- &- &- \\
42$^{*}$ &ILTJ132110.00+362622.5 &0.4 &0.5 &S-N &25.4 $\pm$ 1.9 &51.8 $\pm$ 3.9 &- &- &0.4 $\pm$ 0.3 &- \\
43$^{*}$ &ILTJ132227.31+613251.7 &1.0 &0.4 &N-S &14.6 $\pm$ 1.1 &56.1 $\pm$ 4.2 &- &- &- &- \\
44$^{*}$ &ILTJ134436.36+291239.4 &1.5 &1.2 &E-W &54.8 $\pm$ 4.1 &5.9 $\pm$ 0.5 &11.9 $\pm$ 0.2 &4.3 $\pm$ 0.1 &2.6 $\pm$ 0.2 &0.49 $\pm$ 0.04 \\
45$^{*}$ &ILTJ134713.30+240613.6 &1.1 &2.0 &S-N &724 $\pm$ 54.3 &42.4 $\pm$ 3.2 &- &- &- &- \\
46$^{\mathsection}$ &ILTJ134722.14+582019.2 &0.6 &3.3 &E-W &9.5 $\pm$ 0.8 &8.1 $\pm$ 0.6 &- &- &- &- \\
47$^{\mathsection}$ &ILTJ135847.29+323549.1 &1.4 &0.3 &S-N &15.7 $\pm$ 1.2 &6.6 $\pm$ 0.6 &- &1.1 $\pm$ 0.1 &- &- \\
48$^{*}$ &ILTJ141926.78+662731.9 &0.6 &1.2 &N-S &13.2 $\pm$ 1 &6.4 $\pm$ 0.5 &- &X &- &- \\
49$^{*}$ &ILTJ142015.30+373414.4 &0.7 &0.1 &S-N &33.2 $\pm$ 2.5 &11.9 $\pm$ 0.9 &- &- &0.8 $\pm$ 0.2 &- \\
50$^{\mathsection}$ &ILTJ142334.04+655838.0 &1.3 &1.5 &S-N &57.9 $\pm$ 4.4 &57.3 $\pm$ 4.3 &- &X &- &- \\
51$^{*}$ &ILTJ142712.95+441720.3 &0.6 &0.7 &N-S &12.4 $\pm$ 1 &5.3 $\pm$ 0.5 &3.1 $\pm$ 0.2 &- &- &- \\
52$^{*}$ &ILTJ143431.12+652025.1 &2.0 &0.6 &E-W &83.8 $\pm$ 6.3 &6.2 $\pm$ 0.5 &- &X &- &- \\
53$^{*}$ &ILTJ143532.42+302757.0 &1.3 &1.6 &W-E &97 $\pm$ 7.3 &185.4 $\pm$ 13.9 &1.2 $\pm$ 0.1 &2.5 $\pm$ 0.1 &1.7 $\pm$ 0.3 & $\alpha^{1400}_{144}$ = -0.32 $\pm$ 0.04 \\
"&" &" &" &" &" &" &" &" &" & $\alpha^{3000}_{1400}$ = 0.51 $\pm$ 0.24 \\
54$^{*}$ &ILTJ144609.52+445422.5 &0.4 &1.5 &N-S &224.3 $\pm$ 16.8 &14.3 $\pm$ 1.1 &1.4 $\pm$ 0.1 &0.6 $\pm$ 0.2 &0.8 $\pm$ 0.2 &0.22 $\pm$ 0.18 \\
55$^{*}$ &ILTJ150302.98+572403.3 &- &- &S-N &- &- &- &- &- &- \\
56$^{*}$ &ILTJ150547.75+392528.9 &- &- &S-N &- &- &0.9 $\pm$ 0.2 &0.9 $\pm$ 0.1 &1.1 $\pm$ 0.2 &-0.02 $\pm$ 0.04 \\
57$^{\mathsection}$ &ILTJ152527.43+413215.5 &- &- &N-S &- &- &2.2 $\pm$ 0.2 &- &- &- \\
58$^{\mathsection}$ &ILTJ152756.90+531715.6 &3.5 &0.9 &N-S &11.7 $\pm$ 0.9 &57 $\pm$ 4.3 &- &- &3.9 $\pm$ 0.2 &- \\
59$^{\mathsection}$ &ILTJ152949.12+491557.2 &2.7 &3.1 &N-S &28.8 $\pm$ 2.2 &53.4 $\pm$ 4 &- &- &1 $\pm$ 0.3 &- \\
60$^{*}$ &ILTJ153659.30+310539.5 &2.1 &0.6 &W-E &213.8 $\pm$ 16.1 &95.4 $\pm$ 7.2 &4.7 $\pm$ 0.5 &3.2 $\pm$ 0.1 &2.3 $\pm$ 0.2 &0.22 $\pm$ 0.05 \\
61$^{\mathsection}$ &ILTJ153724.28+425952.5 &1.0 &0.9 &N-S &25.1 $\pm$ 1.9 &25.3 $\pm$ 1.9 &- &0.8 $\pm$ 0.1 &- &- \\
62$^{*}$ &ILTJ154952.31+555120.7 &0.6 &1.0 &E-W &37.7 $\pm$ 2.9 &15.6 $\pm$ 1.2 &2 $\pm$ 0.2 &- &1.5 $\pm$ 0.2 &0.1 $\pm$ 0.06 \\
63$^{*}$ &ILTJ155054.46+473449.3 &0.6 &0.9 &E-W &63 $\pm$ 4.8 &28.7 $\pm$ 2.2 &11.5 $\pm$ 0.2 &6.2 $\pm$ 0.1 &2.7 $\pm$ 0.2 &0.42 $\pm$ 0.16 \\
64$^{\mathsection}$ &ILTJ155436.64+533430.1 &2.1 &1.2 &S-N &18.7 $\pm$ 1.4 &2.6 $\pm$ 0.2 &2.7 $\pm$ 0.4 &1.5 $\pm$ 0.1 &1.4 $\pm$ 0.2 &0.06 $\pm$ 0.24 \\
65$^{\mathsection}$ &ILTJ155805.81+451958.7 &- &- &N-S &- &- &9.7 $\pm$ 0.1 &13.4 $\pm$ 0.2 &14.9 $\pm$ 0.4 &-0.14 $\pm$ 0.01 \\
66$^{\mathsection}$ &ILTJ161010.04+455552.7 &- &- &S-N &- &- &- &- &- &- \\
67$^{\mathsection}$ &ILTJ163415.49+492846.4 &0.2 &1.0 &E-W &85.2 $\pm$ 6.4 &27.5 $\pm$ 2.1 &6.3 $\pm$ 0.3 &3.6 $\pm$ 0.1 &3.8 $\pm$ 0.2 &0.18 $\pm$ 0.06 \\
68$^{*}$ &ILTJ175253.19+372351.1 &1.2 &1.3 &S-N &25.3 $\pm$ 2 &42.4 $\pm$ 3.2 &- &X &- &- \\
69$^{*}$ &ILTJ222719.14+221437.8 &0.7 &1.5 &S-N &111.8 $\pm$ 8.4 &120.1 $\pm$ 9 &- &X &- &- \\
70$^{*}$ &ILTJ230341.84+253002.7 &0.2 &0.4 &S-N &124.3 $\pm$ 9.4 &21.8 $\pm$ 1.7 &4.6 $\pm$ 0.4 &X &2.2 $\pm$ 0.2 &0.25 $\pm$ 0.04 \\
71$^{*}$ &ILTJ231133.28+325119.3 &0.4 &0.1 &E-W &285.9 $\pm$ 21.5 &40.6 $\pm$ 3.1 &3.9 $\pm$ 0.3 &X &7.4 $\pm$ 0.2 &-0.21 $\pm$ 0.03 \\
72$^{*}$ &ILTJ231421.57+192339.3 &- &- &E-W &- &- &- &X &61.1 $\pm$ 0.5 &- \\
73$^{\mathsection}$ &ILTJ232036.32+290056.0 &2.3 &1.6 &W-E &25.4 $\pm$ 2 &3.1 $\pm$ 0.3 &1.5 $\pm$ 0.3 &X &- &- \\
74$^{\mathsection}$ &ILTJ232642.14+184221.9 &- &- &S-N &- &- &- &X &- &- \\
75$^{\mathsection}$ &ILTJ234901.19+231733.0 &0.7 &0.5 &N-S &55.8 $\pm$ 4.2 &62.6 $\pm$ 4.7 &- &X &- &- \\
76$^{\mathsection}$ &ILTJ235751.02+332610.1 &0.5 &0.8 &S-N &10.6 $\pm$ 0.9 &3.4 $\pm$ 0.3 &1 $\pm$ 0.3 &X &- &- \\
\hline \\
1$^{f}$ &ILTJ004145.91+322453.1 &- &- &W-E &- &- &- &X &- &- \\
2$^{a}$ &ILTJ084525.79+522916.4 &0.9 &1.4 &E-W &238.7 $\pm$ 17.9 &68.9 $\pm$ 5.2 &- &3.4 $\pm$ 0.1 &2.9 $\pm$ 0.1 &0.2 $\pm$ 0.08 \\
3$^{c}$ &ILTJ105133.87+514449.9 &- &- &S-N &- &- &- &- &- &- \\
4$^{b}$ &ILTJ105742.51+510558.5 &- &- &E-W &- &- &- &- &- &- \\
5$^{b}$ &ILTJ110457.23+480903.4 &- &- &N-S &- &- &1.6 $\pm$ 0.1 &1.1 $\pm$ 0.2 &- &0.16 $\pm$ 0.07 \\
6$^{b}$ &ILTJ110613.66+485748.3 &0.7 &1.3 &N-S &15.5 $\pm$ 1.2 &2.8 $\pm$ 0.3 &- &- &- &- \\
7$^{c}$ &ILTJ112218.57+555037.0 &- &- &S-N &- &- &- &- &1.1 $\pm$ 0.3 &- \\
8$^{b}$ &ILTJ112425.26+554614.1 &0.8 &0.9 &N-S &15.1 $\pm$ 1.2 &17.2 $\pm$ 1.3 &- &- &- &- \\
9$^{g}$ &ILTJ114722.34+350106.0 &- &- &E-W &- &- &213.8 $\pm$ 8 &615.4 $\pm$ 1.3 &187.2 $\pm$ 0.6 &- \\
10$^{c}$ &ILTJ115528.85+485101.3 &1.9 &0.5 &S-N &273.1 $\pm$ 20.5 &314.1 $\pm$ 23.6 &- &- &3.6 $\pm$ 0.6 &- \\
11$^{c}$ &ILTJ120500.71+475843.6 &- &- &S-N &- &- &- &- &- &- \\
12$^{c}$ &ILTJ121502.39+474640.8 &- &- &E-W &- &- &- &- &- &- \\
13$^{b}$ &ILTJ121859.61+505249.8 &- &- &N-S &- &- &- &- &- &- \\
14$^{c}$ &ILTJ122542.02+520006.1 &1.7 &0.2 &E-W &58.5 $\pm$ 4.4 &43.4 $\pm$ 3.3 &- &1.1 $\pm$ 0.1 &1.5 $\pm$ 0.3 &-0.43 $\pm$ 0.28 \\
15$^{c}$ &ILTJ123005.45+491515.9 &0.9 &1.2 &W-E &107.8 $\pm$ 8.1 &7.4 $\pm$ 0.6 &- &- &- &- \\
16$^{h}$ &ILTJ124237.53+383802.6 &- &- &W-E &- &- &- &1.1 $\pm$ 0.1 &1.6 $\pm$ 0.3 &-0.46 $\pm$ 0.27 \\
17$^{c}$ &ILTJ124548.71+563109.3 &0.7 &1.3 &S-N &15.5 $\pm$ 1.2 &2.8 $\pm$ 0.3 &- &- &- &- \\
18$^{e}$ &ILTJ130126.66+510458.1 &- &- &E-W &- &- &3.8 $\pm$ 0.6 &2.4 $\pm$ 0.2 &1.7 $\pm$ 0.2 &- \\
19$^{c}$ &ILTJ130354.23+464233.5 &- &- &W-E &- &- &1.3 $\pm$ 0.2 &0.8 $\pm$ 0.1 &0.9 $\pm$ 0.1 &- \\
20$^{b}$ &ILTJ134311.00+555940.1 &- &- &N-S &- &- &3 $\pm$ 0.5 &- &4 $\pm$ 0.2 &-0.09 $\pm$ 0.06 \\
21$^{d}$ &ILTJ140719.34+513157.1 &1.7 &0.2 &W-E &58.5 $\pm$ 4.4 &43.4 $\pm$ 3.3 &- &- &- &- \\
22$^{b}$ &ILTJ140719.30+550605.3 &0.9 &1.2 &E-W &107.8 $\pm$ 8.1 &7.4 $\pm$ 0.6 &- &- &- &- \\
23$^{b}$ &ILTJ141504.73+463429.1 &1.6 &1.4 &E-W &120.4 $\pm$ 9.1 &28 $\pm$ 2.1 &0.6 $\pm$ 0.1 &- &- &- \\
24$^{j}$ &ILTJ142335.95+330238.9 &0.5 &3.8 &N-S &191.6 $\pm$ 14.4 &38.7 $\pm$ 2.9 &- &- &1.6 $\pm$ 0.2 &- \\
25$^{j}$ &ILTJ143652.33+341636.7 &2.2 &0.1 &N-S &57.1 $\pm$ 4.3 &10.4 $\pm$ 0.8 &- &- &- &- \\
26$^{c}$ &ILTJ143735.71+514434.8 &1.6 &1.4 &E-W &120.4 $\pm$ 9.1 &28 $\pm$ 2.1 &- &1.5 $\pm$ 0.2 &1.2 $\pm$ 0.2 &0.29 $\pm$ 0.22 \\
27$^{h}$ &ILTJ145302.50+330918.1 &0.8 &1.1 &S-N &8.7 $\pm$ 0.7 &67.7 $\pm$ 5.1 &- &2 $\pm$ 0.1 &7.7 $\pm$ 0.3 &-1.77 $\pm$ 0.1 \\
28$^{c}$ &ILTJ145609.66+482011.2 &0.8 &1.9 &S-N &96.6 $\pm$ 7.3 &11.7 $\pm$ 0.9 &- &- &- &- \\
29$^{c}$ &ILTJ145641.16+484940.5 &1.4 &1.9 &W-E &60.3 $\pm$ 4.6 &37.4 $\pm$ 2.8 &- &- &- &- \\
30$^{a}$ &ILTJ160811.30+325427.2 &0.8 &1.9 &E-W &96.6 $\pm$ 7.3 &11.7 $\pm$ 0.9 &- &3.9 $\pm$ 0.1 &4.6 $\pm$ 0.2 &-0.21 $\pm$ 0.07 \\
31$^{i}$ &ILTJ161242.66+431314.8 &0.9 &0.5 &N-S &723 $\pm$ 54.2 &14.6 $\pm$ 1.1 &- &- &- &- \\
32$^{i}$ &ILTJ161601.20+482535.9 &0.7 &0.6 &N-S &469.3 $\pm$ 35.2 &3.7 $\pm$ 0.3 &2.1 $\pm$ 0.6 &1.4 $\pm$ 0.1 &2.4 $\pm$ 0.2 &0.02 $\pm$ 0.19 \\
33$^{d}$ &ILTJ162754.57+290624.3 &1.4 &1.9 &N-S &60.3 $\pm$ 4.6 &37.4 $\pm$ 2.8 &- &- &- &- \\
34$^{d}$ &ILTJ170517.88+394025.2 &- &- &S-N &- &- &- &- &- &- \\
35$^{d}$ &ILTJ170625.53+434037.4 &- &- &W-E &- &- &- &- &- &- \\

\hline

\end{longtable}
\end{small}

%%%%%%%%%%%%%%%%%%%%%%%%%%%%%%%%%%%%%%%%%%%%%%

\setlength{\tabcolsep}{3.7pt}
\begin{table*}
\captionsetup{width=\textwidth}
\caption{The Right Ascension (RA) and Declination (Dec) of G-DDRGs hosted by the brightest cluster galaxies (BCGs) are listed in columns (2) and (3), respectively. Column (4) provides the cluster names. Column (5) lists redshifts, which are predominantly spectroscopic, except those marked with $^{\dag}$, which are photometric. Column (6) lists the r-band magnitude (r$_{\rm mag}$) of the BCG-DDRGs. The radius $\rm r_{500}$ (column 7) is the radius within which the mean density is 500 times the critical density of the Universe, and $\rm M_{500}$ (column 10) is the mass within this radius. Column (8) shows $\rm R_{L*500}$, the cluster richness parameter, and column (9) lists $\rm N_{500}$, the number of galaxies within $\rm r_{500}$. References for the data are listed in column (11). For galaxy clusters from \citet{WH2015}, $\rm M_{500}$ is computed using Eq. 17 from their paper. For the cluster J014149.6+274046 (marked with $*$), $\rm M_{500}$  is from \citet{Gao2020}.}
\label{tab:clusters}
\centering
\begin{tabular}{l c c c c c c c c c c}
\hline\hline
  SrNo & RA & Dec & Cluster Name  & $z$ & r$_{\rm mag}$  & $\rm r_{500}$ & $\rm R_{L*500}$ & $\rm N_{500}$ & $\rm M_{500}$ & Reference\\

  & (HMS) & (DMS) &  &  &  & (Mpc) &  &  & ($10^{14}$ $\rm M_{\odot}$) & \\
  (1) & (2)& (3)  & (4)& (5)& (6)& (7)& (8)& (9)& (10)& (11)\\
  \hline 
  1& 00 03 45.91 & $+$31 39 24.24 & J000345.9+313924 & 0.5548 & 19.9 & 0.66 & 30.91 & 6 & 1.74 & \citep{WH2015} \\
  2&08 15 38.42 & $+$49 56 00.30 & J081538.4+495601 & 0.5993 & 19.5 & 0.76 & 39.82 & 11 & 2.28 & "\\
  3&08 31 10.57 & $+$44 31 54.22 & J083110.6+443154 & 0.3705 & 18.8 & 0.65 & 20.91 & 10 & 1.14 & "\\
  4&08 32 01.85 & $+$39 55 45.99 & J083201.8+395546 & 0.3216 & 18.9 & 0.73 & 25.56 & 11 & 1.41 & "\\
  5&12 19 00.76 & $+$50 52 54.41 & J121900.8+505255 & 0.3851 & 18.3 & 0.69 & 26.54 & 12 & 1.47 & "\\
  6&13 14 58.77 & $+$38 27 08.7 & J131458.8+382709 & 0.6430 & 21.0 & 0.53 & 20.95 & 7 & 1.14 & "\\
  7&13 16 06.72 & $+$27 19 27.9 & J131606.7+271928 & 0.2497 & 17.5 & 0.73 & 23.63 & 15 & 1.30 & "\\
  8&13 21 10.07 & $+$36 26 27.43 & J132110.1+362627 & 0.2665 & 17.4 & 0.85 & 41.68 & 22 & 2.40 & "\\
  9&14 20 14.04 & $+$37 34 06.52 & J142014.0+373407 & 0.5813 & 20.6 & 0.59 & 19.84 & 7 & 1.07 & "\\
  10&14 53 02.86 & $+$33 08 42.40 & J145302.9+330842 & 0.2481 & 17.7 & 0.71 & 18.68 & 10 & 1.01 & "\\
  11&16 08 10.80 & $+$32 54 18.9 & J160810.9+325419 & 0.3038 & 17.6 & 0.72 & 30.14 & 12 & 1.69 & "\\
  12&23 14 19.63 & $+$19 23 25.93 & J231419.6+192326 & 0.0924 & 15.2 & 0.65 & 16.4 & 8 & 0.88 & "\\
  13&01 41 49.60 & $+$27 40 45.7 & J014149.6+274046 & 0.2702$^{\dag}$ & 17.1 & - & - & - & 1.38$^{*}$ & \citep{WHY2018} \\
  14&15 37 23.92 & $+$42 59 56.8 & J153723.8+425959 & 1.4015$^{\dag}$ & 25.1 & 0.52 & - & 14 & 1.36 & \citep{WH2021} \\
15& 16 34 15.05 & $+$49 28 42.8 & 061323 &0.517 & 9.7 & -  & - & - &- &  \citep{Szabo2011} \\
\hline
%\tablecomments{}
\end{tabular}
\end{table*}

\newpage

%%%%%%%%%%%%%%%%%%%%%%%

\section{Radio images of giant DDRGs} \label{sec:appendixB}

\begin{figure*}
  \includegraphics[scale=0.10]{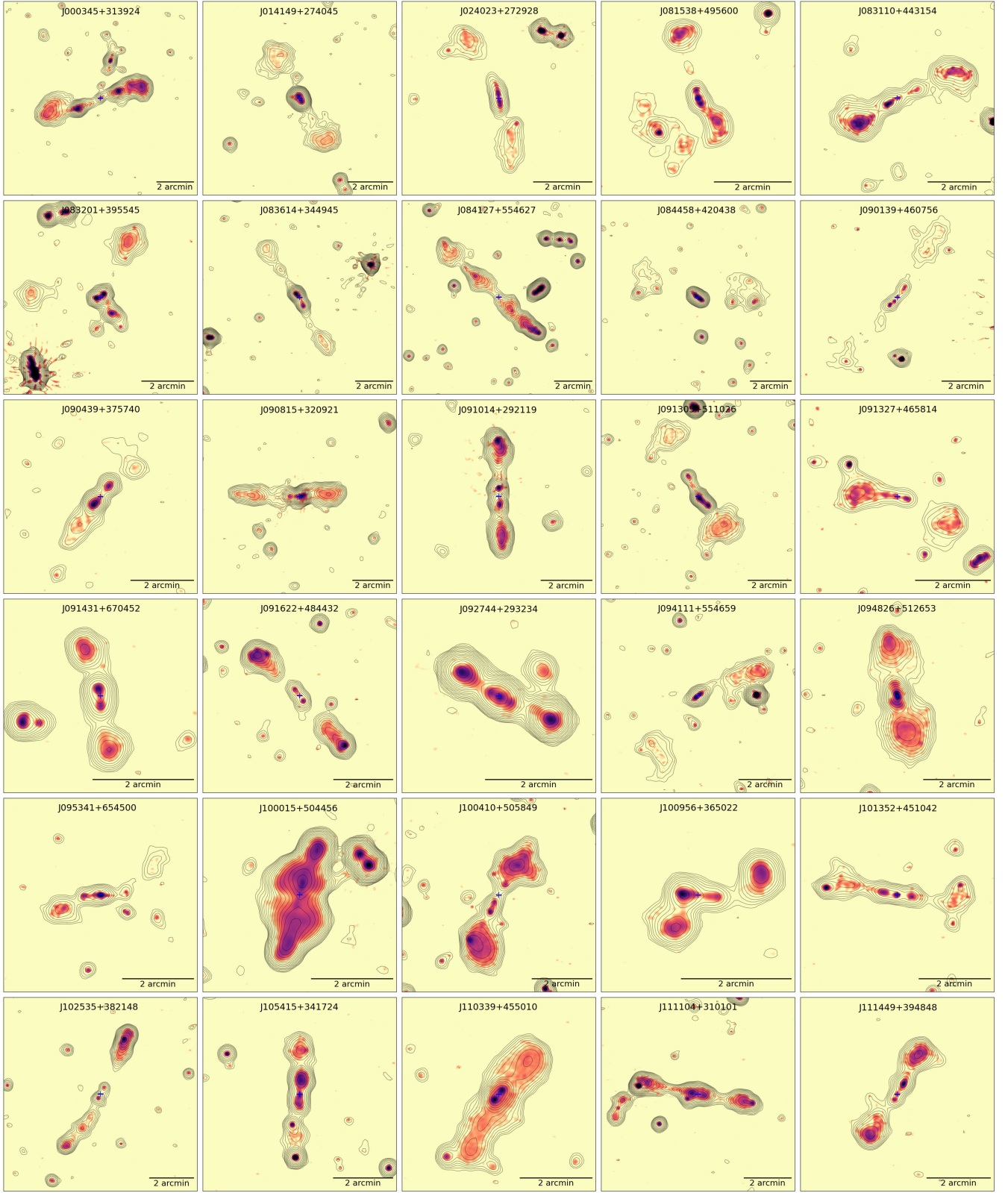}
  \caption{\label{fig:Mosaic1} LoTSS DR2 144 MHz radio maps of giant DDRGs listed in Tab.~\ref{tab:main} (1 to 30). In this series of montages, we first present the images of the G-DDRGs discovered by us followed by the ones known in the literature with different colour schemes. The colour and contours from LoTSS have angular resolutions of 6\arcsec~ and 20\arcsec respectively, showing emission above 3$\sigma$. The RMS or $\sigma$ is typically $\sim$\,~80\mujybeam. Each successive contour is $\sqrt{2}$ times the previous contour level.}
\end{figure*}

\begin{figure*}
  \includegraphics[scale=0.10]{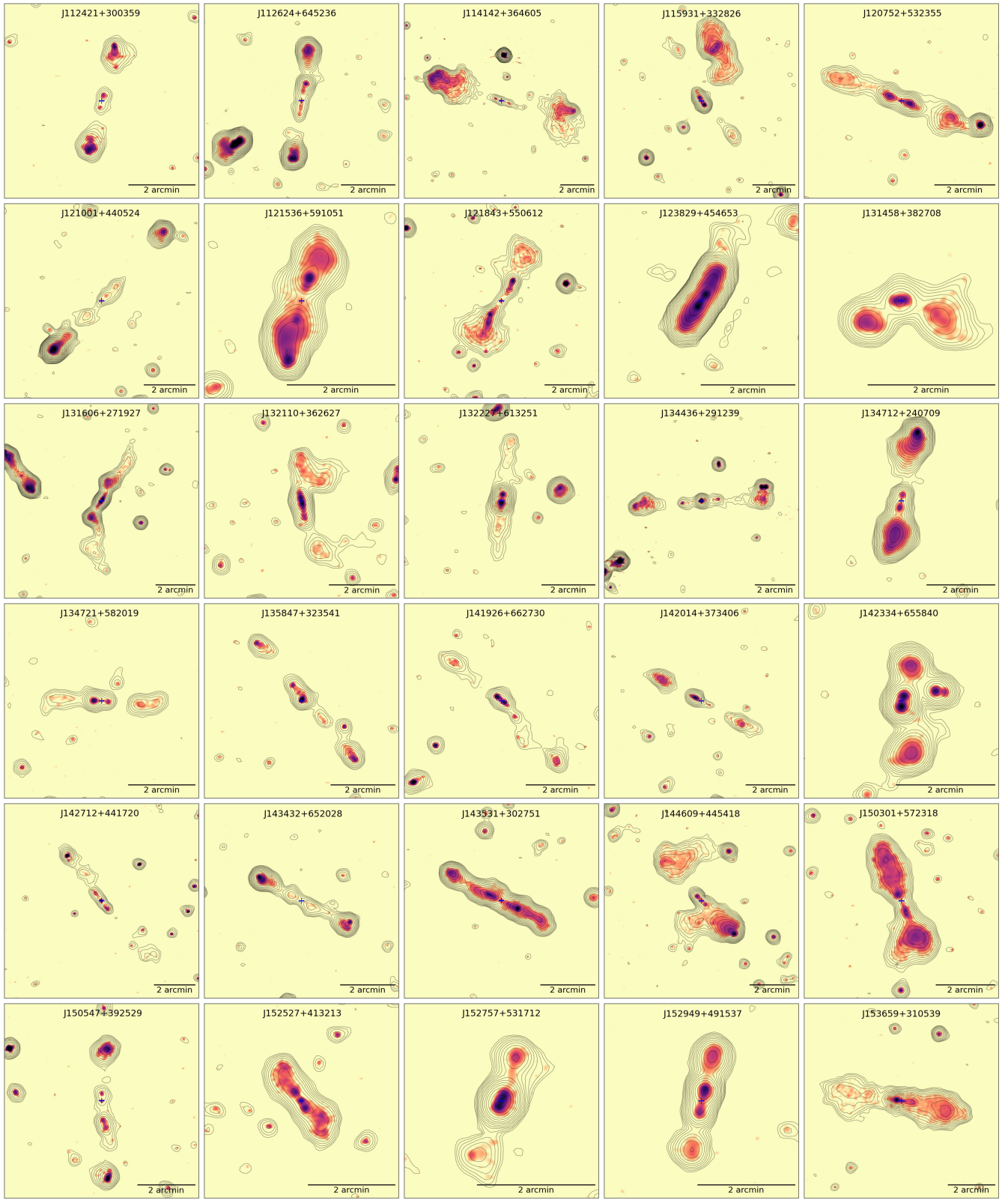}
  \caption{\label{fig:Mosaic2} Same as Fig.~\ref{fig:Mosaic1} but for sources 31 to 60 from Tab.~\ref{tab:main}.}
\end{figure*}

\begin{figure*}
  \includegraphics[scale=0.10]{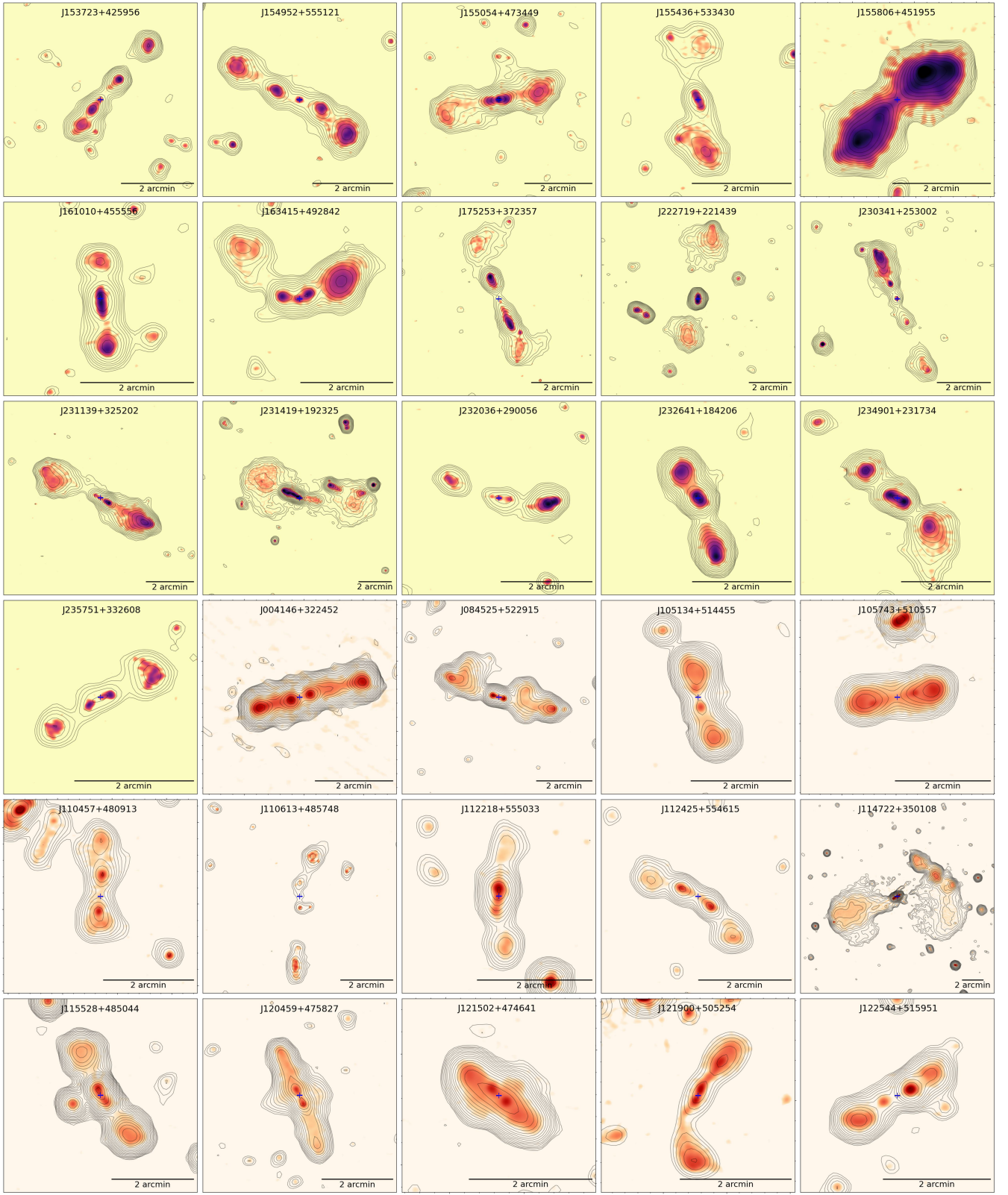}
  \caption{\label{fig:Mosaic3} Same as Fig.~\ref{fig:Mosaic1} but for sources 61 to 76 from the upper half of Tab.~\ref{tab:main} and sources 1 to 14 (in different colour scheme) from the lower half of the Tab.~\ref{tab:main} which represent known G-DDRGs in our sample (for more details see Sec.~\ref{sec:sample}).}
\end{figure*}

\begin{figure*}
  \includegraphics[scale=0.10]{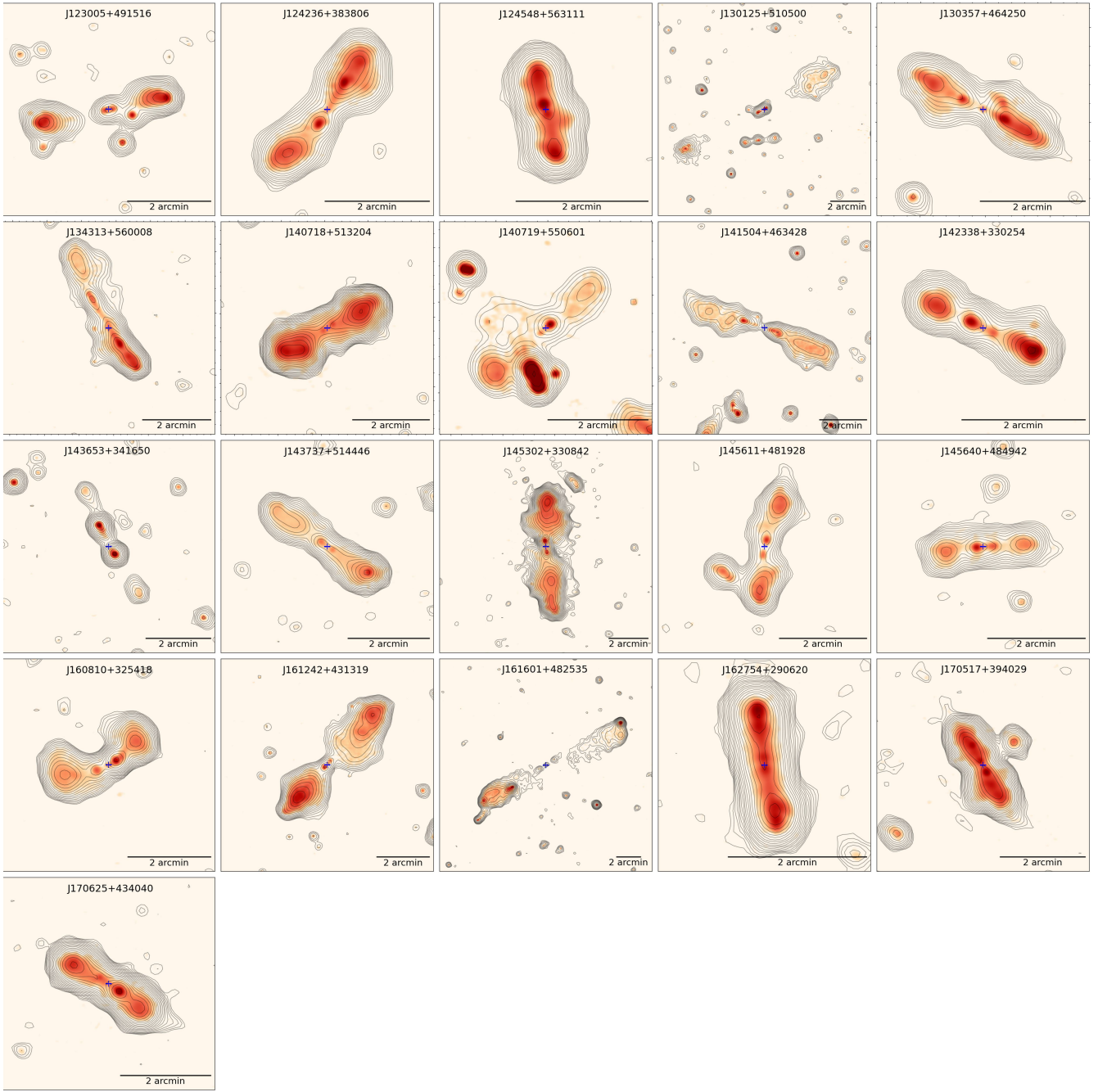}
  \caption{\label{fig:Mosaic4} Same as Fig.~\ref{fig:Mosaic1} but for sources 15 to 35 from the lower half of the Tab.~\ref{tab:main} which represent known G-DDRGs in our sample.}
\end{figure*}

\end{document}